\documentclass[10pt,a4paper]{article}

\usepackage{mathtext}
\usepackage[T1]{fontenc}
\usepackage[dvips]{graphics}
\usepackage{amsmath}
\usepackage{amssymb}
\usepackage{textcomp}         

\usepackage{float}            
\usepackage{wrapfig}          
\usepackage{natbib}          

\begin{document}

\title{Resilience of Volatility}

\author{Sergey S. Stepanov
\footnote{Altus Assets Activities R\&D; {\tt steps@altus.ua}}
\footnote{Dnepropetrovsk Center for Fundamental Research, Ukraine. }
}

\maketitle\thispagestyle{empty}

\begin{abstract}
The problem of non-stationarity in financial markets is 
discussed and related to the dynamic nature of price volatility.
A new measure is proposed for estimation of the current asset 
volatility. A simple and illustrative explanation is suggested 
of the emergence of significant serial autocorrelations in 
volatility and squared returns. 
It is shown that when non-stationarity is eliminated, the
autocorrelations substantially reduce and become statistically
insignificant.
The causes of non-Gaussian nature of the probability of returns 
distribution are considered.
For both stock and currency markets data samples, it is shown 
that removing the non-stationary component substantially reduces 
the kurtosis of distribution, bringing it closer to the Gaussian one.
A statistical criterion is proposed for controlling the
degree of smoothing of the empirical values of volatility.
The hypothesis of smooth, non-stochastic nature of volatility is put
forward, and possible causes of volatility shifts are discussed.
\end{abstract}



\section{Introduction}

Non-stationarity is arguably the most characteristic feature of financial 
markets. It is generally accepted that statistical parameters of price 
dynamics vary with time. This fact is unpleasant both for researchers and 
practitioners, because any discovered regularities and elaborated
trading systems quickly lose their relevance as time passes. The best solution
to the problem of non-stationarity would be to include it into a
probabilistic model of market operation.

One of the most important characteristics of returns of a
financial instrument is its volatility. There is no doubt that
volatility varies over time, and this phenomenon is the subject of
voluminous literature, for example \cite{Schwert:1989}, as well as a
more recent collection in \cite{Shephard:2005}. There are
'quiet' periods of market behavior and periods of increased
volatility. One can say that volatility characterizes the market 
'temperature', the degree of its emotional tension. Forecasting 
future values of volatility is extremely important; it plays a crucial 
role, among other issues, in determining the pricing of
options and assessing the risk for portfolio investors (see 
\cite{Poon:2003} for an extensive review).
Understanding the causes and nature of non-stationary volatility
would also lead to a deeper insight into the essence of the 
financial market processes. Various models were suggested, encompassing
such diverse fields as theory of chaos applied by \cite{Hsieh:1991}, 
and multi-agent systems studied by \cite{Watanabe:2008}. 
This task became especially relevant in recent years, during the unfolding 
financial disturbances, as well as dramatic events of Internet bubble 
(relevant discussion can be found in \cite{Battalio:2006}).

The term {\em volatility}
comprises at least four different meanings: 1) the emotional characteristic 
of the market; 2) sample mean square deviation of logarithmic returns; 3) the
'true' unobservable variance of the underlying distribution of returns; and 
4) the implied volatility in option contracts. 
In this paper, we use the term {\em volatility} in the second and third 
senses, which are usually referred to as realized and latent volatility,
respectively. We refer the reader to reviews by \cite{McAleer:2008}
and by \cite{Shephard:2008} for good overview of the field. 
The choice of robust volatility estimator is important for producing
correct inferences from the available data (see \cite{Broto:2004}).
One of the questions we focus on in our research is, which 
choice of estimator of sample volatility leads to
a minimum error for a certain {\em model} of a random process.
'True' volatility is, of course, non-observable, and
the question of its nature is further complicated by the
non-stationary nature of the markets.

The generally accepted approach is to consider the volatility as a
stochastic variable (see, for example, \cite{Andersen:1998},
\cite{Shephard:2005} and the collection in \cite{Forecasting:2007}). 
One of the chief
motivations for this is the presence of high autocorrelations in
volatility and squared returns, as discussed in \cite{Engle:2001}. 
Compared to the
near-zero autocorrelations in logarithmic returns, the
detection of such long-memory pattern creates striking impression.

In the probabilistic models with variable volatility, the price 
$x(t)$ random process is described either by discrete or continuous equations,
parameters of which are random variables. In this context, GARCH($p,q$) model 
first introduced by \cite{Engle:1982} gained wide popularity, as well 
as its various generalizations (see \cite{Engle:2002}). In this case,
the timeline is divided into finite time intervals (lags) of
duration $\tau$, and then only 'closing' prices of these
intervals are considered $x_k = x(k\cdot\tau)$, 
where $k = 1,2...$ is an integer. 
{\em Logarithmic returns} $r_k =\ln(x_k/x_{k-1})$ are
independent random variables, with variable volatility $\sigma_k$,
the square of which linearly depends on the previous squared
returns and volatilities:
\begin{equation}
  r_k = \sigma_k ~\varepsilon_k,
~~~~~~\sigma^2_k = \alpha_0 + \sum^p_{i=1} \alpha_i
     ~\sigma^2_{k-i}  + \sum^q_{i=1} \beta_i ~r^2_{k-i}.
\end{equation}
Here and below $\varepsilon_k$ is uncorrelated
normalized random (i.i.d.) process with zero mean and unit variance:
$\overline{\varepsilon_i} = 0$, ~$\overline{\varepsilon^2_i} = 1$, 
~$\overline{\varepsilon_i\cdot\varepsilon_j} = 0$. A line
over a symbol, as usual, denotes the average of all the possible
realizations of $\varepsilon_i$.

In the continuous framework, the stochastic Ito's equation is
widely used for both price and volatility dynamics. The price is 
modeled (for an example, see a paper by \cite{Alizadeh:2002}) by the 
ordinary logarithmic walk, and volatility is described by the
Ornstein--Uhlenbeck equation:
\begin{equation}
\frac{dx}{x} = \mu dt + \sigma(t)\delta W_1 ,~~~~~~ d\ln\sigma =
\beta\cdot(\alpha - \ln\sigma) ~ dt + \gamma\delta W_2,
\end{equation}
where $\delta W_1$, $\delta W_2$ are uncorrelated Wiener
variables $\delta W = \varepsilon\sqrt{dt}$. Indeed, the term
'stochastic' volatility is usually reserved for this class of
models, although we here use in a somewhat broader sense, to include 
the GARCH-type models.

Sometimes, both in the discrete and the continuous models, one or
more 'hidden' stochastic variables are introduced, and volatility 
is considered as a function of such variables. 
Other, sometimes rather sophisticated approaches, exist in the
literature (see \cite{Eraker:2004} as an example). What
unites them all is the probabilistic description of the local
dynamics of volatility (either discrete or continuous).

There is an extensive body of empirical research devoted to  
testing of predictive power of GARCH-type stochastic models over the 
last twenty years, surveyed in \cite{Canina:1993}, 
\cite{Forecasting:2007},
as well as the discussion of correct methodology for forecast 
estimation \cite{Andersen:1998}. 
In general, certain skepticism regarding the predictive capabilities 
of such models is present in ongoing research. Recently,
certain considerations were expressed that explain the persistence of
autocorrelations of positively determined variables as the result of 
their non-stationarity; \cite{Mikosch:2004}, 
\cite{Granger:1999} and \cite{Diebold:2001} are just a few 
examples of related research.

The effect of non-stationarity is also directly related to the problem
of searching for the probability distribution of returns of financial 
instrument. 
It is well known that this distribution is non-Gaussian; it has heavy
tails and, consequently, manifests high kurtosis and high probability of 
excessively large or small returns. 
Starting with the seminal work by \cite{Mandelbrot:1963}, this 
fact has gradually become a standard in financial engineering (see 
\cite{Jondeau:2007} for a modern view upon the subject). 
However, most approaches to constructing the probability
distribution of random variables implicitly suppose their
stationarity, which we do not observe at real financial markets.

The idea that non-stationarity in the random process can cause the
non-Gaussian behavior of returns distribution goes back as far as
the classical work by \cite{Fama:1965}; it was therein
tested and was not confirmed. Nevertheless, the question about the
type of distribution and the effect of non-stationarity requires
further careful consideration.

In this paper we provide the arguments in support of the hypothesis that
volatility $\sigma(t)$ is a {\em smooth}, rather than stochastic,
function of time. The explanation of origin of high long-term 
autocorrelations and the non-Gaussian nature of returns distribution will 
be given. 
Our hypothesis also implies that the volatility manifests
the property of resilience: under the impact of irregular, relatively rare 
and completely unpredictable shocks to the market, it gradually deforms;
after such influences cease to act, the relaxation process takes over and
volatility gradually decreases.

The remainder of the paper is organized as follows.
First, we discuss a new measure of volatility and
demonstrate its effectiveness. After that, the empirical stylized facts 
of autocorrelations associated with the volatility
are listed, and a simple non-stationary model, in which such
properties naturally arise, is proposed. 
A very clear graphic representation is provided for the mechanism of 
appearance of autocorrelations, 
and a simple mathematical formalism for performing the necessary calculations
is proposed. 

The evidence that such a mechanism is actually realized in
financial markets is provided by calculation of the
autocorrelation function (ACF) for two modifications of original
series; namely, the autocorrelations are vanishing for both the
first differences of volatility of consecutive days, and for the
residual series obtained by elimination of its smooth part
$\sigma(t)$. The empirical tests of these facts are carried out
utilizing sample data of both stock market and exchange rate
dynamics.

Next, we show that normalizing the returns series by $\sigma(t)$
leads to significant reduction in kurtosis of distribution,
in some cases restoring it to the normal form. Statistical
criteria for controlling the degree of data smoothing are
elaborated. We consider the arguments concerning the local
constancy of 'true' volatility. In Conclusion, a number of
inferences about the possible properties of the dynamics of
volatility are formulated. Various technical details are compiled 
in self-contained Appendices, which complement and detail the 
calculations presented in the body of the paper.

\section{Measurement of volatility}

Historical prices for various financial instruments are usually
available as discrete time series, with a certain period of time
(lag) between consecutive points. Most widely available are data
with daily lags, hourly lags can be observed less frequently, and
minute lags are even more seldom. In addition to the closing price
$C_t$ (the latest value within a lag), other commonly utilized
parameters are: opening quotes $O_t$ (first price of a lag),
maximal $H_t$ and minimal $L_t$ price values. By means of these
four price points one can construct three independent {\it
relative} values, which we will call the {\it basis} of a lag.
\begin{figure}[H]
\begin{minipage}{.55\textwidth}
\centering
\includegraphics{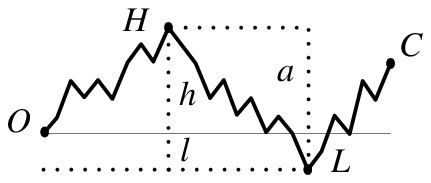} 
\caption{Characteristics of volatility}\label{fig:1}
\end{minipage}\hspace*{1cm}
\begin{minipage}{.35\textwidth}
\begin{eqnarray}
h_t&=&H_t-O_t \nonumber\\
l_t&=&O_t-L_t \\
r_t&=&C_t-O_t \nonumber
\end{eqnarray}
\end{minipage} 
\end{figure}
The {\it height} $h$ of price ascent and the {\it depth} of price
descent $l$ are both positive values. Out of these measures the
{\it amplitude} of price range $a = h + l$ can be defined (see, for example 
\cite{Parkinson:1980} and \cite{Garman:1980} for early 
examples of its use). 
The asset return $r$ can be both positive and negative.

If one considers the models of additive Wiener random walk
$dx=\mu~dt+\sigma~\delta W$, the values $\{O_t,H_t,L_t,C_t\}$ are
asset prices. For the logarithmic random walk
$dx/x=\tilde{\mu}dt+\tilde{\sigma}\delta W$ they are logarithms of price values
$\ln x$. Thus, in the latter case, for example, the range $a_t$
would be equal to the logarithm of the ratio of maximum price to
minimum price $a_t=\ln(H_t/L_t)$, the corresponding $r_t$ equal to
the logarithmic return $r_t=\ln(C_t/O_t)$, and so on.

We define volatility $\sigma$ of a lag with duration $T$ as an
average of asset return $r$ deviation from the mean over a
sufficiently large number of lags:
$\sigma^2=\left<(r-\bar{r})^2\right>$. If volatility $\sigma$ is
constant, the values of positive entities $\{h,l,|r|,a\}$ in
certain sense serve as its measure. The higher is the market
volatility, the more probable are their high values. In
particular, in absence of drift ($\mu = 0$), their men values are
proportional to volatility:
$\bar{a}=1.596\cdot\sigma,~\bar{h}=\bar{l}=\overline{|r|}=0.798\cdot\sigma$
(see \ref{appxA}).

\begin{figure}[H]
\centering
\includegraphics{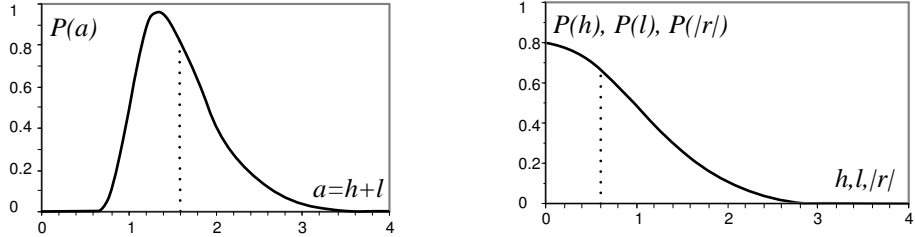}
\caption{Probability densities for the basis components}\label{fig:2}
\end{figure}

However, the informational content of each parameter, and of their
possible combinations, varies. The distributions of the
probability density for $P(a)$ and $P(h)$, $P(l)$, $P(|r|)$, in
the case of Wiener process, are plotted in Fig.~\ref{fig:2} (dotted lines
mark the distributions' mean for $\sigma=1$).
As we see, among the basis components $\{a,h,l,|r|\}$ only the range 
$a$ has a
sufficiently narrow maximum around the mean value. The density of
probability of the other three values are strictly decreasing
functions, which allows for $h$, $l$ and $|r|$ to take, with high 
probability, values close to zero. The range $a$,
on the contrary, avoids going to zero, the probability that $a<0.75\sigma$
being as low as 0.002. Indeed, it often happens
that the market closes with a near-zero change in price $|r|\sim 0$, 
while its
volatility during the day was significant. In general, the narrower the
distribution of probability for volatility measure, the better is this 
measure.
For some positively determined value $v$, the relative degree of
distribution narrowness can be characterized by a ratio
$\sigma_v/\bar{v}$, where $\sigma^2_v=\overline{(v-\bar{v})^2}$ is
mean squared deviation from the mean $\bar{v}$. For the range we
have $\sigma_a/\bar(a)=0.30$, which signifies more than twice as
narrow distribution peak than, for instance, for the height
($\sigma_{h}/\bar{h}=0.76$). A natural question arises: is there a
combination of the basis values $f(h,l,r)$ that has a narrower
distribution than the price range $a$? This topic is the subject
of extensive research (see e.g.
\cite{Parkinson:1980},\cite{Garman:1980},
\cite{Rogers:1991},\cite{Rogers:2008},
\cite{Yang:2000}).

\begin{figure}[h]
\begin{minipage}[c]{.55\textwidth}
\centering
\includegraphics{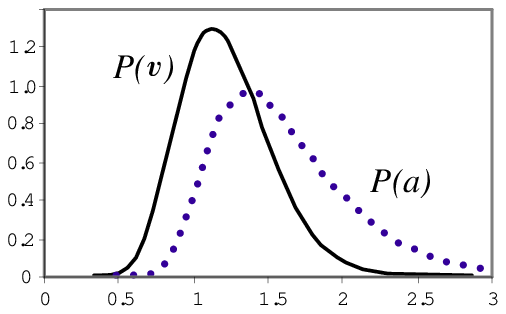}
\end{minipage}\hspace*{1cm}
\begin{minipage}[c]{.35\textwidth}
$$
 \bar{v} = \frac{3}{\sqrt{2\pi}}\cdot\sigma,
$$
$$
 \overline{v^2} = \left(4\ln 2 -\frac{5}{4}\right) \cdot \sigma^2.
$$
\end{minipage} 
\caption{Probability
densities $P(v)$, $P(a)$ and mean values for $v$ and $v^2$}\label{fig:3}
\end{figure}

In the present article we define a simple, but efficient, modification 
of the price range, which is motivated as follows. If the price dynamics 
within the lag is
accompanied by a significant trend $|r|\neq 0$ (whether it is going
up or down), the volatility may appear lower than for the
same price range, but in the absence of trend ($|r|=0$).
Therefore there are good reasons to decrease the value of
the range, as a measure of volatility, in the case when $|r|$ is large. 
We achieve this by introducing the following volatility estimator, 
which we call {\it modified price range} (see \ref{appxB}):
\begin{equation}
          v = a - \frac{|r|}{2}.
\end{equation}
Its statistical parameters -- mean ($av$), standard deviation ($si$), skewness
($as$) and its kurtosis ($ex$) for $\sigma=1$ are listed in Table~\ref{tbl:1}.
\begin{table}[H]
\caption{Statistical parameters of probability distributions for $|r|$, $a$ and $v$.}
\label{tbl:1}
\centering
\small
\begin{tabular}{c|rrrr|r}
        & $av$          & $si$   & $as$        & $ex$     & $si/av$\\
\hline
$|r|$   &  0.798        & 0.603    &  1.00      & 0.87     & 0.76 \\
$a$   &  1.596        & 0.476    &  0.97      & 1.24     & 0.30 \\
$v$   &  1.197        & 0.300    &  0.53      & 0.26     & 0.25 \\
\end{tabular}
\end{table}
One can see that the relative width of the distribution of modified
range $\sigma_v/\bar{v}=0.25$, which is better than that of simple
range $a$. The statistical parameters also show that the
distribution for $v$ is more symmetrical around the maximum and
has a lower kurtosis than $a$. The form of
distribution for $P(v)$ together with $P(a)$ (dotted line) are plotted
in Fig.~\ref{fig:3}, and there we also provide the expressions for the average 
$v$ and its square for the case of the Brownian walk.
Thus, the modified price range provides a better measure of
volatility than the simple range, and significantly better than
absolute logarithmic returns. In \ref{appxB}, we compare the
modified range with several other ways of volatility measurement
utilized by other authors. Providing for the same or lower error
of volatility estimation, the measure $v$ has a significantly more
simple definition, and is unbiased for the small number of lags,
so we will use it extensively throughout this paper.

\section{Intraday volatility}

We shall demonstrate the effectiveness of modified amplitude of
range on the available realized volatility data. Here we consider
15-minute quotes at the Forex market for the period from 2004 to
2008 for EURUSD currency pair. We shall make them aggregated into
daily points, calculating, beside minimum and maximum meanings,
intraday volatility basing on logarithmic returns of 15-minute
lags:
\begin{equation}\label{sigma_sum_r2}
   \sigma^2 = \frac{n}{n-1} \sum^n_{i=1} (r_i-\bar{r})^2.
\end{equation}
During a day, we have $n=96=4\cdot24$ 15-minute lags. Multiplier
$n$ in (\ref{sigma_sum_r2}) turns 15-minute volatility into the
daily value. The evolution of {\it intraday volatility} is given in
the Fig.~\ref{fig:4} (data for 1250 trading days, excluding weekends and 
major holidays):
\begin{figure}[H]
\centering
\includegraphics{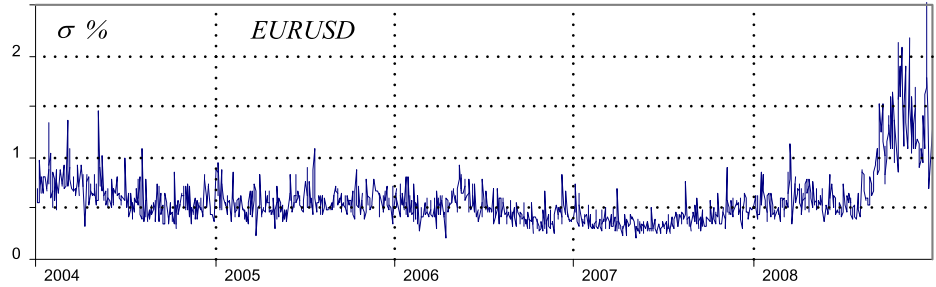}
\caption{Intraday volatility of EUR/USD rate}\label{fig:4}
\end{figure}
One can observe that since the fall of 2008, volatility of the
currency market, as well as of other financial markets, has
increased dramatically, due to the worsening financial crisis.
However, even in the pre-crisis period, volatility has a clear-cut
non-stationary component.

\begin{figure}[h]
\centering
\includegraphics{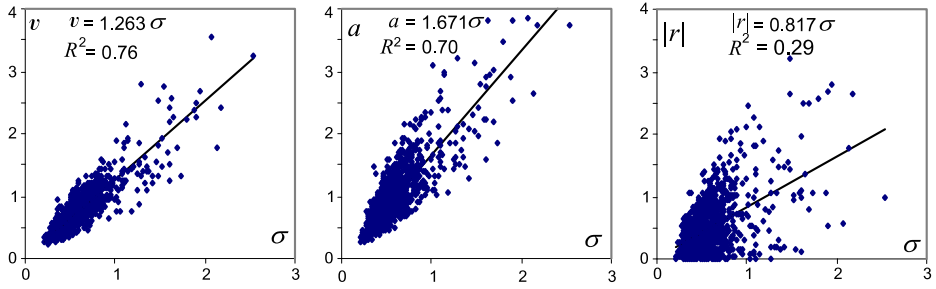}
\caption{Dependencies
of $v(\sigma)$, $a(\sigma)$ and $|r|(\sigma)$ }\label{fig:5}
\end{figure}

It is natural to assume that realized volatility obtained from a
sample of $n=96$ characterizes the 'true' volatility better than
does a daily basis of three values $\{h,l,r\}$ (see
\cite{Barndorff:2002}, \cite{Biais:2005},
\cite{Andersen:2003}) even though there are various
high-frequency effects that one has to take into account
(discussed in detail by \cite{Madhavan:2000},
\cite{Bandi:2006}, and \cite{Bollerslev:1993}). To
find a more robust measure of volatility, based on the basis, one
should look for a value stronger correlated with the intraday
volatility. Let us chart the scatter plots of dependence of daily
values of $v_t$, $a_t$ and $|r_t|$ on intraday volatility
$\sigma_t$ (EUR/USD for period 2004-2008, Fig.~\ref{fig:5}).
It can be easily seen that $v_t$ and $a_t$ are substantially more
correlated with $\sigma_t$, than with $|r_t|$. The transition from
logarithmic range $a$ to modified range $v$ makes the correlation more
pronounced, but the difference is not significant.

Similar results are observed for other currencies. The slope of
regression lines $v_t/\sigma_t$ and $a_t/\sigma_t$ for six
currency pairs are given in Table~\ref{tbl:2}.
In each case the error of linear approximation for $v$ is lower
than for $a$, and significantly lower than for $|r|$.
\begin{table}[H]
\centering
\small
\caption{Slope of regression lines $v_t/\sigma_t$ and $a_t/\sigma_t$ for six pairs of currencies}
\label{tbl:2}
\begin{tabular}{r|rrrrrr|rrrrrrrrrrrr}
                             &  {\small eurusd} & {\small gbpusd} & {\small usdchf}  & {\small usdjpy}   & {\small usdcad} & {\small audusd}  & average \\
\hline
$\left<v/\sigma\right>$      &  1.263           &          1.260  &          1.289   &          1.251    &  1.241          &   1.243          & 1.258       \\
$\left<a/\sigma\right>$      &  1.671           &          1.665  &          1.692   &          1.640    &  1.621          &   1.660          & 1.658 \\
$\left<|r|/\sigma\right>$    &  0.817           &          0.809  &          0.807   &          0.776    &  0.761          &   0.834          & 0.801 \\
\end{tabular}
\end{table}
Despite the noticeable variation, the values of
$v/\sigma$,$a/\sigma$ and $|r|/\sigma$ are close to their
theoretical values for Wiener random walk, 1.197, 1.596 and 0.798, 
respectively. Nevertheless, we must keep in mind that, for example,
the expression $v/\sigma=3/\sqrt(2\pi)$ holds {\it only} for the
Brownian random walk with {\it normal} distribution of returns.
In reality, this condition is not fully satisfied, so the ratio 
$v/\sigma$ may be equal to some constant different from
$3/\sqrt(2\pi)$, and its exact value we will discuss below.

\begin{figure}[H]
\centering
\includegraphics{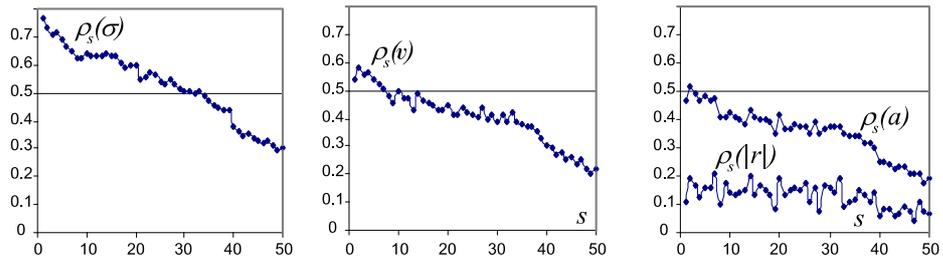}
\caption{Correlograms of volatility for EURUSD}\label{fig:6}
\end{figure}

Another indication of significance of modified price range are 
autocorrelation coefficients that will now be the object of
our interest:

\begin{equation}
   \rho_s(v) ~=~ cor(v_t, v_{t-s}) ~=~ \frac{\left<(v_t-\bar{v})(v_{t-s}-\bar{v})\right>}{\sigma^2_v},
\end{equation}
where the averaging is carried out for all the observed values of
$v_t=v_1,..,v_n$. For interdaily rates of EUR/USD (2004-2008) we
obtain (as shown in Fig.~\ref{fig:6}) the autocorrelation charts as a function of 
shift (in days) parameter $s$.
As can be observed from Figure~\ref{fig:6}, autocorrelations of intraday
volatility $\rho_1(\sigma)=0.77$ are the highest, followed by
modified price ranges $\rho_1(v)=0.54$, then by
simple range $\rho_1(a)=0.47$, and the weakest correlations are those 
of absolute logarithmic returns $\rho_1(|r|)=0.11$.

High autocorrelations appear for a variety of financial instruments and 
are quite an intriguing fact (\cite{Cont:2001} provides the
list of other so-called stylized facts, as well as an excellent compilation
of references to relevant research). In contrast, the first 
autocorrelation coefficient of EUR/USD rate returns is equal 
to $\rho_1(r)=-0.02$,
which corresponds to the absence of correlation, if one takes into account 
that $2\sigma$ rule gives an error band of 0.06 (for 1250 trading days). This
unpredictability of the market returns is one of manifestations of its
market effectiveness.

However, the situation is quite different for absolute returns, and
even more so for volatilities, which have slowly decaying long-range ACF 
function. Basing on this fact a huge number of
stochastic models have been constructed, which claim the ability to
predict the future values of volatility (see \cite{Shephard:2008} for
a recent review). The majority of these models have {\it empirical} 
nature, and do not explain the {\it causes} of high autocorrelations. 
One of our tasks in the present paper will be to
propose such an explanation


\section{Empirical features of autocorrelations}

We now extend our analysis by outlining a number of features of autocorrelation 
coefficients pertinent for volatility.

\begin{quote}
1. {\it Autocorrelations decay monotonically and very slowly}.
\end{quote}
This is a well-known result (see \cite{Ding:1993},
\cite{Breidt:1998}). A number of papers were devoted to attempts
on determining the functional dependence of autocorrelation
coefficients from the shift parameter $s$. Usually, autocorrelations
are approximated by a power law $s^{-\mu}$, where the parameter
$\mu$ turns out to be small.

\begin{quote}
2. {\it The longer is the time interval, the higher are autocorrelations}.
\end{quote}
Let us consider the behavior of ACF for daily modified range 
$v=a-|r|/2$ for S\&P500 stock
index for the period from 2001 to 2006. We split this interval
into two three-year periods, namely, from 2001 to 2003 and from 2004 to 2006. 
During the first one there were $n=752$ trading days, while during the second -
$n=755$. We calculate autocorrelation coefficients separately for each
period, as well as the autocorrelation of combined data series.

\begin{figure}[H]
\centering
\includegraphics{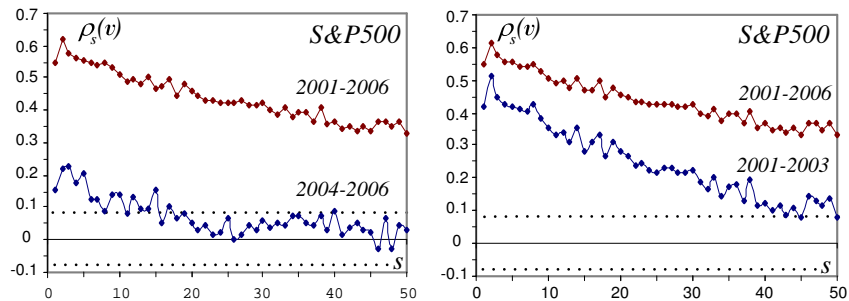}
\caption{Correlograms
of S\&P500 for different time periods}\label{fig:7}
\end{figure}

The resulting autocorrelograms are represented in Fig.~\ref{fig:7}
(the combined ACF is repeated on both plots).
As can be noticed, the summarized correlogram goes above the
correlograms of each period. However, this behavior is not
observed for any asset in any circumstances, and the conditions
that are required for this to occur will be clarified during
further discussion below.

Here and below the dotted horizontal lines in the correlograms
mark the double standard error band $\pm2/\sqrt(n)$, where $n$ 
is the number of points involved into the calculation.
Table~\ref{tbl:3} shows the main statistical
parameters of daily logarithmic returns of S\&P500 index for
different periods.
\begin{table}[h]
\centering
\small
\caption{Main statistical parameters of daily logarithmic returns of S\&P500 index}
\label{tbl:3}
\begin{tabular}{rrrrrrrrrrrrrrrrrrr}
Period      & $n$     &  $\overline{r}$& $\sigma$  &~~~~$as$ & $ex$& ~~~~~~$p_0, \%$  &  $p_1, \%$  & $\rho_1(v)$\\
\hline
2004-2006   &  755     &   0.032   &  0.659       & -0.02  & 0.25   & 55.9  &  69.4  & 0.16 \\
2001-2003   &  752     &  -0.023   &  1.376       & 0.20   & 1.27   & 48.9  &  71.4  & 0.42 \\
2001-2006   & 1507     &   0.005   &  1.078       & 0.15   & 2.84   & 52.4  &  75.7  & 0.55 \\
\end{tabular}
\end{table}
In addition to the mean ($\overline{r}$), daily volatility
$\sigma$, skewness($as$) and kurtosis($ex$), we also present here the percentage of
positive returns $p_0 = p(r>0)$ and a share of returns falling within 
one sigma of the mean: $p_1=p(|r-\bar{r}|<\sigma)$.

Table~\ref{tbl:3} illustrates the fact that when the market is calm
(2004-2006: $\sigma=0.659\%$), the distribution of asset returns is 
close to normal ($ex=0.25$). However, the normality deteriorates 
significantly after we extend the time interval under consideration. 
Simultaneously the autocorrelation of volatilities starts to increase 
$\rho_1(v)=cor(v_t, v_{t-1})$.

\begin{table}[h]
\centering
\small
\caption{Volatility autocorrelation coefficients for EUR/USD exchange rate, with and without 2008Q4 data}
\label{tbl:3x}
\begin{tabular}{rrrrrrrrrrrrrrrrrrr}
Period               & $n$     &  $\rho_1(\sigma)$ & $\rho_1(v)$ & $\rho_1(a)$  &  $\rho_1(|r|)$ \\
\hline
2004Q1~..~2008Q4   &  1302    &  0.80        &       0.54       & 0.47 & 0.11 \\
2004Q1~..~2008Q3   &  1215    &  0.51        &       0.25       & 0.16 & 0.01 \\
\end{tabular}
\end{table}

A similar situation can be observed in the foreign exchange market.
Discarding the data from recessionary fourth quarter of 2008 reduces
significantly the autocorrelation coefficients of data series
related to the volatility of EUR/USD pair, as illustrated in
Table~\ref{tbl:3x}.
We note that dropping the 2008Q4 data reduces the number of days for which the
autocorrelation coefficients are calculated by merely 7\%.

\begin{quote}
3. {\it Scatter plot of volatility has a 'comet-like' shape}.
\end{quote}

Let us consider the scatter plots for modified range parameter
$\{v_{t-1}, v_t\}$, illustrating the 'existence of memory' of volatility 
for the three periods of $S\&P500$ index, discussed above.
\begin{figure}[ht]
\centering
\includegraphics{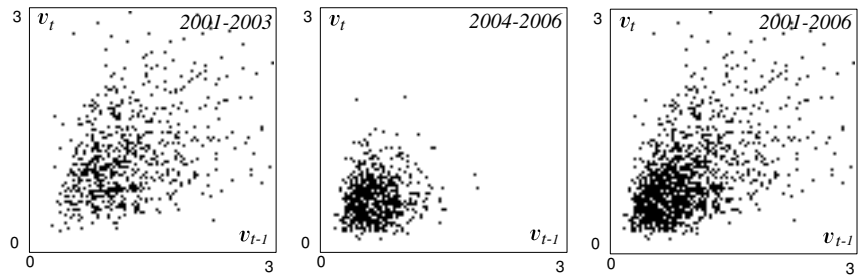}
\caption{Dependence diagrams $a_t(a_{t-1})$, $S\&P500$}\label{fig:8}
\end{figure}
As can be seen from Fig.~\ref{fig:8}, data points fill the region of a
distinctive 'comet-like' shape, its tail fanning out into the positive values
of both axes. Naturally, the higher the autocorrelation coefficients
are, the more distinctive is the form of the dot cloud.

\begin{figure}[H]
\centering
\includegraphics{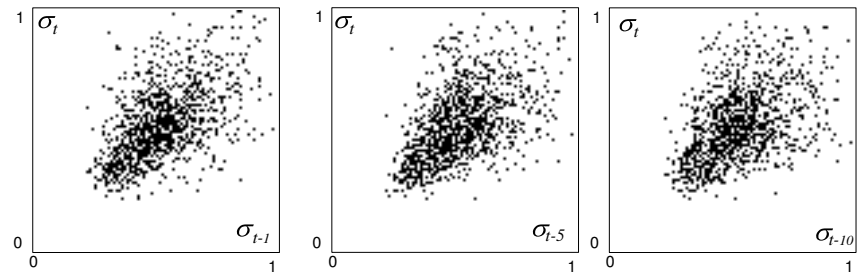}
\caption{Dependencies
$\sigma_t(\sigma_{t-s})$, $s=1,5,10$, EURUSD }\label{fig:9}
\end{figure}

The shape of region $\sigma_t=f(\sigma_{t-s})$ is completely independent
of the shift $s$ and the utilized volatility measure.
For EUR/USD currency pair over 2004-2008 period, we
have the scatter plots of intraday volatilities, obtained
from 15-minute lags, are presented in Fig.~\ref{fig:9}.
There, three values of the shift are presented: one day ($s=1$),
one week ($s=5$), and two weeks ($s=10$). It can be seen that the 
form of 'comet-like' shape doesn't change qualitatively, but rather 
spreads out gradually along with the decrease of autocorrelation coefficient.

\section{When autocorrelations do not decay}

Actually, slowly decreasing autocorrelation coefficients as a function of
the shift parameter, ought to be a cause for alert. There
are very simple models that exhibit similar long-correlations effects 
without employing the notion of stochastic volatility (see
\cite{Granger:1999} for one ingenuous example).

Let us consider, for example, an ordinary logarithmic walk:
\begin{equation}
   \frac{dx}{x} = \mu~dt+\sigma~\delta W.
\end{equation}
and simulate 20 years (5000=20$\cdot$250 trading
days) of price evolution; volatility is defined as constant equal to 
$\sigma_1=1\%$ for the first 10 years, and changes to another constant
value of $\sigma_2=2\%$ in the second half of the period.
Wiener's process is represented as $\delta W=\varepsilon\sqrt{dt}$, 
where $\varepsilon$ is normally distributed random variable with zero 
mean and unit variance. We choose
one second $dt=1/(24\cdot 60\cdot 60)$ as a small time interval $dt$.

\begin{figure}[H]
\centering
\includegraphics{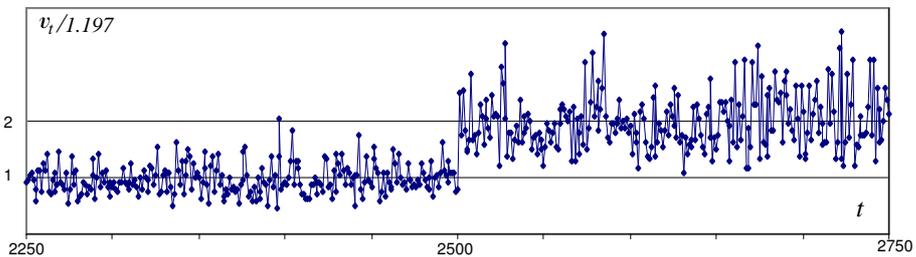}
\caption{Two years of
random walk around the 'switch' in volatility}\label{fig:10}
\end{figure}

\begin{figure}[H]
\centering
\includegraphics{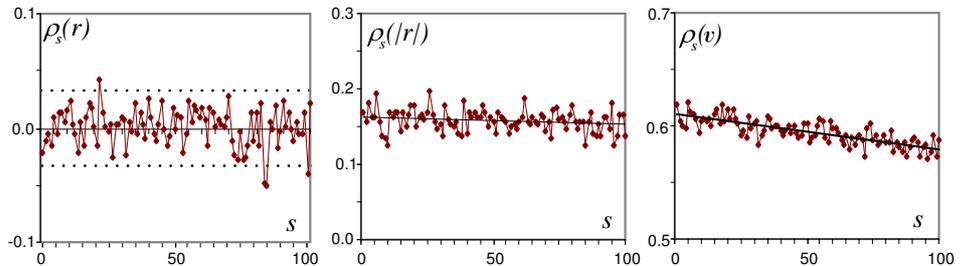}
\caption{Autocorrelations of returns, 
absolute returns and modified price range.}\label{fig:11}
\end{figure}

The dynamics of daily values of the modified price range
$v_t=a_t-|r_t|/2$ during 'critical' 10th and 11th years has the
shape plotted in Fig.~\ref{fig:10} (where time is in 'days').
Such data series with a one-time shock non-stationarity exhibits
noticeable autocorrelation coefficients for the absolute returns
(plotted in the second panel of Fig.~\ref{fig:11}), and even higher autocorrelations
for the price range (third panel).
The decay of ACF is very slow with the increase of
shift parameter $s$. In contrast to $|r_t|$ and $v_t$,
the correlations of price returns $r_t$ (the first plot above)
lie within two standard errors, and thus are practically absent.

Therefore, correlation regularities arise in the considered toy model, 
despite the statistical
independence of the two consecutive days. 
We stress that not only the returns $r$ are independent,
but so are the absolute returns $|r|$, and amplitudes of price
$v$. If volatility were constant for the whole modeled period, all
the correlograms $\rho_s(|r|)$ and $\rho_s(v)$ would be equal to
zero. It is when we introduce non-stationarity that the picture is
qualitatively changed.

The cause of this effect can be easily understood. Fig.~\ref{fig:11}
contains three scatter plots that represent values of logarithmic returns,
their absolute values and modified price ranges of two consecutive days
during the first decade of evolution with constant volatility
$\sigma=1\%$.
\begin{figure}[ht]
\centering
\includegraphics{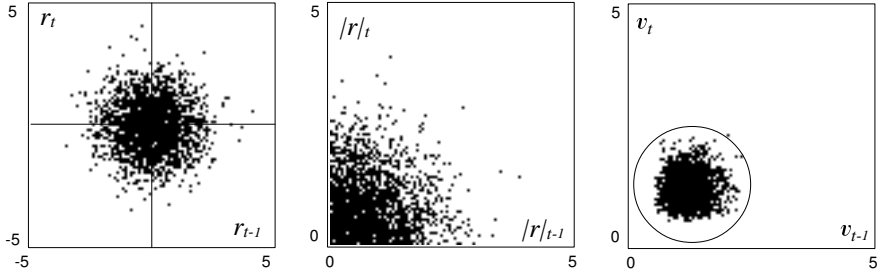}
\caption{First decade of the model}\label{fig:12}
\end{figure}
In the first plot, the dots form an almost symmetrical cloud,
and the correlation is evidently equal to zero. In the second and
third plots, there is symmetry is reduced, in agreement with the
corresponding symmetry features of the probability densities
$P(|r|)$ and $P(v)$. However, due to the independence of
consecutive days, the correlation coefficient is equal to zero.
For example, if $x=v_t$, and $y=v_{t-1}$, the independence means
that the joint density of distribution is equal to the product of
probability densities $P(x,y)=P(x)\cdot P(y)$. Therefore, for {\it
any} distribution the covariance will be equal to zero:
$\overline{(x-\bar{x})(y-\bar{y})}=0$.

It is important to emphasize the fact that for the returns $r_t$
the center
of the data cloud is located at the origin of coordinates, whereas for
the positively determined values $|r_t|$ and $v_t$ it is displaced to
the right and up to the region of positive values.

\begin{figure}[ht]
\centering
\includegraphics{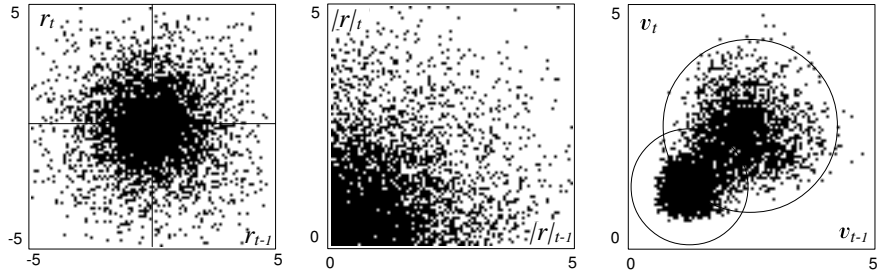}
\caption{The complete data of the model}\label{fig:13}
\end{figure}

Now let us add the dots corresponding to the data of the second decade 
to the diagram (see Fig.~\ref{fig:13}).
For logarithmic returns (first diagram) two clouds with the same
center $\bar{r}=0$ overlay. The resulting cloud remains
symmetrical, that is why the autocorrelation is still equal to
zero. In the case of the price range (third diagram),
there are two non-concentric clouds, one of which corresponds to
$\sigma_1=1\%$ and second one to $\sigma_2=2\%$ (we remind
that $\bar{v}=1.197\sigma$). The overlapping area between the
clouds dithers, and a figure of a characteristic comet-like
shape appears as a result (the upper cloud is larger). Using the 
least-squares criterion, one can draw a line through it, the slope 
which will be proportional to the correlation coefficient.

The shape of diagrams do not change if we plot the data for the case 
of two days'
shift $\{v_{t}, v_{t-2}\}$. Indeed,
with the exception of few transitional points around the volatility jump, 
all the data for each decade will still be clustered in its cloud.

The situation with the second diagram for the absolute returns
$\{|r_{t-1}|, |r_t|\}$ is somewhat more complicated. Visually, it is not
qualitatively different from the corresponding one for the
first decade; nevertheless, the non-zero correlation is present.
In order to understand this phenomenon, it is necessary to extend the standard
statistical relations to the case of non-stationary data.

\section{Non-stationary statistics}

Let the distribution parameters of a random variable $x$ vary
smoothly with time. If we calculate the mean of $x$
over a given time interval $T$ without taking the above mentioned
statement into account, we will actually obtain the following expression
for $\bar{x}$:
\begin{equation}
   \bar{x} = \left<\bar{x}(t)\right>=\frac{1}{T}\int\limits^T_0  \bar{x}(t) ~dt,
   ~~~~   where~~~~\bar{x}(t)=\int\limits^\infty_{-\infty} x\cdot P(x,t)~dx.
\end{equation}
In other words, in every fixed moment of time we calculate a {\it
local mean} $\bar{x}(t)$, and then average all such local mean values
over the time interval $T$ (denoted by angle brackets). Similarly,
let us define {\it local variance} as:
\begin{equation}
 \sigma^2(t) = \int\limits^\infty_{-\infty} (x-\bar{x}(t))^2 \cdot P(x,t)~dx
= \overline{x^2}(t) - \overline{x}(t)^2.
\end{equation}
The variance calculated on all dataset will be equal to:
\begin{equation}\label{eq9}
   \sigma^2 ~=~ \overline{(x-\bar{x})^2} =
\bigl<\overline{x^2}(t)\bigr>-\bigl<\overline{x}(t)\bigr>^2 =
\bigl<\sigma(t)^2\bigr> ~+~ \bigl<\overline{x}(t)^2\bigr> - \bigl<\overline{x}(t)\bigr>^2,
\end{equation}
where the angle brackets, as above, denote averaging over time
interval $T$. Thus, $\sigma^2$ is made up of two distinct parts,
namely, it is a sum of weighted local variance $\left<\sigma^2(t)\right>$ 
and time variance of mean (second and third terms in equation (\ref{eq9})).

In the special case of {\it parametric non-stationarity}, $x_t$ 
can be represented as $x_t=\mu(t)+\sigma(t)\cdot
\eta_t$, where $\eta_t$ represents stationary independent
random process with zero mean and unit variance
($\overline{\eta}=0$, $\overline{\eta^2}=1$). The mean value of $x_t$
is equal to $\bigl<\mu(t)\bigr>$, and variance is given by:
$\bigl<\sigma(t)^2\bigr> + \bigl<\mu(t)^2\bigr> - \bigl<\mu(t)\bigr>^2$.

Let us now consider two {\it locally} independent variables $x$
and $y$. Their independence means that the density of joint
probability in any fixed moment of time $t$ decomposes into product
$P(x,y,t)=P(x,t)\cdot P(y,t)$, and
\begin{equation}
  \overline{x\cdot  y}\; (t)  = \overline{x} (t)\cdot \overline{y}(t).
\end{equation}
However, when averaged over all data, these variables cease being
independent. Indeed, the time mean of the product $x\cdot y$:
\begin{equation}
 \overline{x\cdot y} = \frac{1}{T}\int\limits^T_0  \int\limits^\infty_{-\infty} x\cdot y ~P(x,y,t) ~dx dy dt =
\frac{1}{T}\int\limits^T_0 \bar{x}(t) \bar{y}(t) ~dt = \left<\bar{x}(t)\cdot\bar{y}(t)\right>,
\end{equation}
and this expression is not equal to the product of time means:
 $\overline{x} \cdot \overline{y} =
\left<\bar{x}(t)\right>\cdot\left<\bar{y}(t)\right>$.
In general, if local
means $\bar{x}(t)$, $\bar{y}(t)$ are non-zero, the correlation coefficient
is non-zero as well. As we observed from the example of the previous section,
the mean of returns in each decade was equal to zero, that is why the
autocorrelation did not arise for $r$. In contrast, for the
positively determined variables $|r|$ and $a$ the mean is non-zero, and
autocorrelation is present, despite the independence of
two consecutive days. 

Thus, locally independent variables that have similar long-term 
non-stationarity, become dependent when we take into account their
evolution in time. However, such dependence does not have stochastic
nature, but rather 'deterministic', smooth one, related to time
synchronization.

For example, if the non-stationarity of volatility has a shape of
a step-function with equal duration of both periods, 
the mean and variance of the whole dataset are equal to:
\begin{equation}
   \bar{x}=\frac{\bar{x}_1+\bar{x}_2}{2},~~~~~~~~~~~~~\sigma^2=\frac{\sigma^2_{x1}+\sigma^2_{x2}}{2} + \frac{(\bar{x}_1-\bar{x}_2)^2}{4},
\end{equation}
where $x$ replaced either $|r|$ or $v$, and statistical parameters
of the first and second decades are given by $\bar{x}_1$, $\sigma_{x1}$, and
$\bar{x}_2$, $\sigma_{x2}$. If the
shift in the calculation of the autocorrelation coefficient is
small compared to the length of $T$, in the first approximation one 
can neglect the boundary
effects, and assume that $x=v_t$ and
$y=v_(t-1)$ are independent. Their covariance is
equal to:
\begin{equation}
 \overline{x\cdot y}  - \overline{x} \cdot \overline{y}
~=~ \frac{\bar{x}_1\bar{y}_1+\bar{x}_2\bar{y}_2}{2} - \frac{\bar{x}_1+\bar{x}_2}{2}\cdot \frac{\bar{y}_1+\bar{y}_2}{2}
~=~ \frac{(\bar{x}_1-\bar{x}_2)(\bar{y}_1-\bar{y}_2)}{4}.
\end{equation}
As $\bar{x}_i=\bar{y}_i$, we receive for the autocorrelation
coefficient:
\begin{equation}
   cor(x,y) = \frac{(\bar{x}_1-\bar{x}_2)^2}{(\bar{x}_1-\bar{x}_2)^2+2(\sigma^2_{x1}+\sigma^2_{x2})}.
\end{equation}
We can see that such 'correlation' for non-stationary
data appears only for variables with different means. For $v$ and
$|r|$, mean and variance are proportional to volatility of
logarithmic returns $\bar{x}=\alpha \sigma$, $\sigma_x=\beta
\sigma$. For absolute returns, $\alpha=0.795$, $\beta=0.605$,
while for the modified price range $\alpha=1.197$,
$\beta=0.300$. If volatility of the market changes, so do the mean
values of positively determined variables $v$
and $|r|$. For our model, the volatility of logarithmic return
is increased by factor of 2, and the corresponding correlation 
is given by $1/(1+10(\beta/\alpha)^2)$. For absolute returns this
is equal to 0.15, and for the modified range, 0.61. This agrees exactly
with what we have observed in the numerical experiment.

In general case, in order to obtain the autocorrelation
coefficients as function of shift $s$, we should use their
definition as sums. However, the presentation in continuous time
is more compact. Let us assume $T$ and shift $s$ to be continuous
variables. In case of $n$ lags with duration of $\tau$ each we
have $T=n\tau$, and $s=k\tau$, where $k\ll n$. Let us consider a
positively determined variable $\sigma_t$, related to volatility,
which is modulated by a non-stationary component
$\sigma_t=\sigma(t)\cdot \theta_t$, where $\theta_t$ is a
stationary random variable with a unit mean. For example, for the
modified amplitude, $\theta_t=v_t\sqrt{2\pi}/3$. Since random
variables $\theta_t$ at different times are non-correlated and
positively determined, we obtain that $\overline{\theta_t\cdot
\theta}_{t-s}$ is equal to 1 for $s\neq 0$, and to
$\overline{\theta^2}$ for $s=0$. Let us define the covariance for
the case $s\neq 0$ as follows:
\begin{equation}\label{autocov_nontaticionarity}
\gamma_s(\sigma) = \left<\sigma_t\cdot \sigma_{t-s}\right> - \left<\sigma_t\right>^2
= \frac{1}{T-s} \int\limits^T_s \sigma(t)\sigma(t-s) ~dt
- \left[\frac{1}{T} \int\limits^T_0 \sigma(t)~dt\right]^2.
\end{equation}
We note that this is not the only possibility in the case of a finite
sample with duration $T$. In any case, we require that the
covariance(\ref{autocov_nontaticionarity}) is equal to zero if
$\sigma(t)=const$. The variance of a positive variable $\sigma_t$
equals to
$\gamma_0=\overline{\theta^2}\,\left<\sigma^2(t)\right>-\left<\sigma(t)\right>^{2}$.
Accordingly, the autocorrelation coefficient
$\rho_s=\gamma_s/\gamma_0$ allows to find the shift parameter dependence 
for different forms of non-stationarity.

We thus see that autocorrelations of various measures of volatility can
arise due to smooth non-stationarity in data, rather than because
of the stochastic nature of volatility. At this point, a natural
question comes up: does such a mechanism represent the reason why
noticeable autocorrelations of volatility are observed in various
financial markets?

\section{Autocorrelation of differences}

The easiest way to eliminate the relatively smooth
non-stationarities in a time series is to switch to the
differences of the data series. If $v_t$ undergoes a locally
constant drift, autocorrelations for this process are present. If
one considers the differences of two consecutive data points, the
drift is effectively cancelled. Even if the trend in $v_t$ slowly
changes its direction, within the ascending and descending parts
the values of differences change only slightly and become locally
quasi-stationary.

\begin{figure}[h]
\centering
\includegraphics{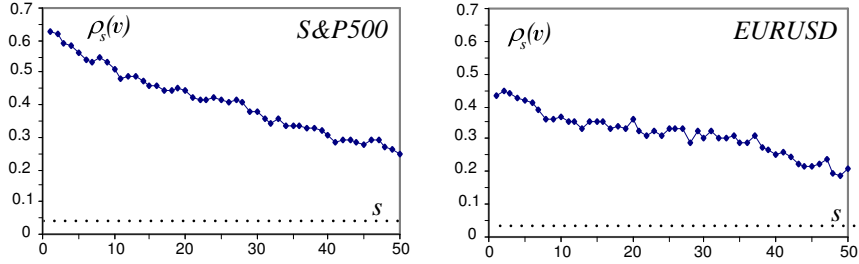}
\caption{S\&P500 and
EUR/USD correlograms of the daily of price range}\label{fig:14}
\end{figure}

Let us shall consider the change in the modified price range:
\begin{equation}
\delta v_t=v_t-v_{t-1}.
\end{equation}
Our data sample is represented by daily statistics on S\&P500
stock index for the period of 1990-2008 (4791 trading days), and
daily data on EURUSD exchange rate (1999-2008, 2495 days,
excluding holidays). Let us start with obtaining the
autocorrelation coefficients of the amplitudes of daily price
range $\rho_s(v)=cor(v_t,v_{t-s})$, with result plotted in
Fig.~\ref{fig:14}. As usual, the coefficients $\rho_s$ are considerably
high; the autocorrelations for S\&P500 index are more significant
than those for EUR/USD exchange rate, and manifest weaker
fluctuations.

\begin{figure}[ht]
\centering
\includegraphics{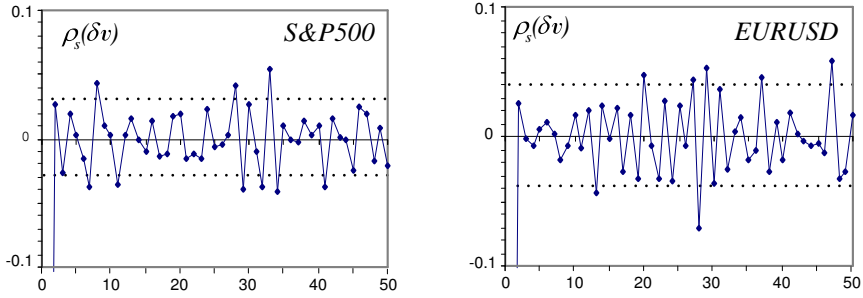}
\caption{S\&P500 and
EURUSD correlograms after for differentiated price range}\label{fig:15}
\end{figure}

Let us now consider the differences the price range of two
consecutive days; we find that for differentiated series, the
autocorrelation $\rho_s(\delta v)$ drops sharply, as can be seen
from Fig.~\ref{fig:15}.
The dissimilarity between these two behaviors is striking. The
second autocorrelation coefficient for S\&P500 index is reduced by
factor of 24, from the value of 0.618 to 0.026. For the EUR/USD
rate the decline is 17-fold -- from 0.449 to 0.027.
Dotted lines in all figures indicate the double standard error,
equal to $0.03 = 2/\sqrt{4791}$ for S\&P500 index and $0.04 =
2/\sqrt{2495}$ for EUR/USD. 
The disappearance of correlation
can be manifestly demonstrated by means of the scatter plots of
consecutive values of $v_t$ and $v_(ts)$.

\begin{figure}[H]
\centering
\includegraphics{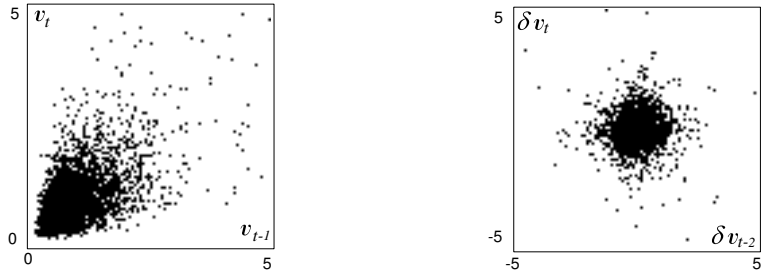}
\caption{First aurocorrelation
for S\&P500 before and after switch to differences}\label{fig:16}
\end{figure}

The two diagrams plotted in Fig.~\ref{fig:16} clearly show the presence of correlations
between $\{v_t, v_{t-1}\}$ of S\&P500 index, and their absence for
$\{\delta v_t, \delta v_{t-2}\}$.
In the left chart dots fill the area with a characteristic
comet-like shape, while in the right one they form a symmetrical
cloud of zero correlation. The similar results, with autocorrelation
coefficients being equal to zero, are also obtained for absolute 
logarithmic
returns $|r_t|$, as well as for other financial instruments.

We note, however, that for the differences $\delta v_t$ a high
negative autocorrelation appears for a shift of one day,
$\rho_1(\delta v)=cor(\delta v_t, \delta v_{t-1})$. In the above
example it is equal to -0.49 for S\&P500 index and -0.53 for
EUR/USD. However, its origin is not due to the stochastic
dynamics of volatility, but rather caused by the overlapping effect. 
We now elucidate it by way of example. Let us assume that
the following simple model governs the price process:
\begin{equation}\label{simple_model}
        v_t = \sigma \cdot \theta_t,
\end{equation}
where $\sigma=const$, and $\theta_t$ are stationary independent
positive random variable that arises because of the errors caused by the
finiteness of the sample that is used for volatility measurement.
In this case, the differences $\delta v_t =
\sigma\cdot(\theta_t-\theta_{t-1})$ have zero mean
$\overline{\delta v_t} = 0$. The first autocorrelation coefficient
equals to
\begin{equation}
\left< \delta v_t \cdot \delta v_{t-1}\right> =\sigma^2 \left<
(\theta_t-\theta_{t-1}) \cdot (\theta_{t-1}-\theta_{t-2})\right> =
- \sigma^2 \cdot \left[~\overline{\theta^2} -
\overline{\theta}^{\,2}~\right] = -\sigma^2\cdot \sigma^2_\theta,
\end{equation}
where $\sigma^2_\theta$ is the variance of $\theta$.
The mean of square arises in the term $-\left< \theta_{t-1}\cdot
\theta_{t-1}\right>= \overline{\theta^2}$, which is the one responsible
for the effect of overlap. In the same way, the variance of difference
$\left<\delta v^2_{t}\right> = 2\sigma^2\sigma^2_\theta$ is
obtained. Thus, first autocorrelation coefficient is exactly
equal to $\rho_1(\delta v)=-0.5$, as we have seen above.
Correlations with shifts of $s>1$ will be equal to zero, because
there is no overlap in this case.

The fact that for the autocorrelations of differences the 
relations $\rho_1(\delta v)=-0.5$ and $\rho_s(\delta v)=0$ ($s>1$) 
hold with a good degree of accuracy corroborates the model 
(\ref{simple_model}). However, if the
parameter $\sigma$ were a constant, there would be no correlation
between consecutive values of volatility $\rho_s( v)=0$ (due to
the independence of $\theta_t$). The correlation may occur, as we have shown
above, as a consequence of {\it gradual} change in $\sigma$
over time. Therefore, actually $\sigma=\sigma(t)$ is a
smooth function of time.

Both for the conclusive clarification of the situation with 
$\rho_1(\delta v)$,
and for the purposes of further research, we
need a method for extracting of the smooth non-stationary
component of volatility.

\section{Filtering smooth non-stationarity}

For the {\em extraction} of slowly varying component in
the process $x_k=x(t_k)$ we will use the Hodrik-Prescott filter
\cite{Hodrick:1997} 
(referred to as HP-filter below). The smooth component 
$s_k$ of the series can be found by way of minimizing the squares of
its deviations from empirical data $x_k$, along with the
requirement of curvature minimality for $s_k$:
\begin{equation}
    \sum^n_{k=1} ( x_k-s_k)^2 + \lambda \cdot \sum^{n-1}_{k=2} (\nabla^2 s_k)^2  = min,
\end{equation}
where the second difference is given by $\nabla^2
s_k=(s_{k+1}-s_k)-(s_{k}-s_{k-1})$. The higher $\lambda$ parameter
is, the more smooth shape $s_k$ one receives as the result. The
value of $\lambda$ can vary in a very wide range, so it is convenient to 
use it's decimal logarithm $\nu$ instead, so that $\lambda = 10^\nu$.

When one deals with heavily noisy data, there is always certain
freedom in the choice of $\lambda$ parameter. If $\lambda$
is small, there is a danger of detecting bogus non-stationarity where it
does not exist. With little smoothing,  $s_k$ component will
follow any local fluctuations, which do not have any relation to
non-stationarity. On the other hand, with strong smoothing we
risk missing important details of the process dynamics that is the focus of our
interest.

Therefore, we need a certain statistical criteria of the degree of
smoothing in order to reduce the possible arbitrariness. As usual,
we will use the random walk as the yardstick.

The mean value of logarithmic returns is equal to the relative
change in price within the time lag $r_t=\ln C_t/O_t$. We 
measure the volatility basing on a smoothed mean of modified
range within a lag
$\sigma(t)=\overline{(a-|r|/2)}\cdot\sqrt{2\pi}/3$. Here, when using the term
'volatility', we always assume volatility of a lag (whether it is minute,
hour, day, etc.).

If the number of discrete price ticks within a lag is
sufficiently large, then regardless of the intra-lag distribution,
logarithmic returns $r_t$ will be uncorrelated {\it Gaussian}
random numbers. Let us smooth their mean value $\bar{r}(t)$
using the HP-filter with different parameters $\lambda$ and
calculate the typical value of $Err\bigl[\bar{r}(t)\bigr]$ for
fluctuations $\bar{r}(t)$ around the average $\bar(r)$ for all
empirical data:
\begin{equation}
   Err\bigl[\bar{r}(t)\bigr]= \sqrt{\left<(\bar{r}(t)-\bar{r})^2\right>}.
\end{equation}
Similarly, we determine the error of calculation of smoothed
volatility of a lag. Our numerical simulations show that these errors,
with a good degree of accuracy, decrease as the 
$\lambda$ parameter grows, as follows:
\begin{equation}\label{hp_err_aver}
   Err\bigl[\bar{r}(t)\bigr] \approx \frac{0.50~\sigma}{\lambda^{1/8}},~~~~~~~~~~~~~~~~
   Err\bigl[\sigma(t)\bigr] \approx \frac{0.15~\sigma}{\lambda^{1/8}},
\end{equation}
and manifest no noticeable dependency on the number of empirical points
$n$. Moreover, the errors do not depend on the type of
distribution (for a discrete model of random walk). The rather small power 
exponent of 
1/8 clarifies the reason why one needs to vary 
$\lambda$ parameter over a wide range of values.

The expressions (\ref{hp_err_aver}) define a typical corridor of
oscillations for the smoothed variables $\bar{r}(t)$ and $\sigma(t)$, which
are fluctuations and are not statistically significant {\it for
constant volatility}. Therefore, we use them as criteria of
statistical significance,
at least for the sections of data where $\sigma(t)$ is approximately
constant.

\begin{figure}[H]
\centering
\includegraphics{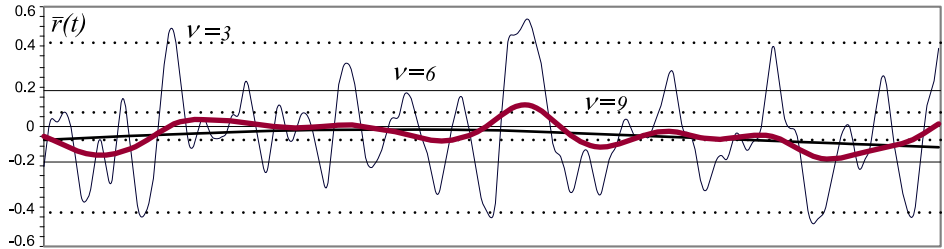}
\caption{Smoothed mean of Gaussian noise}\label{fig:17}
\end{figure}

Let us consider a typical example of a numerical simulation ($\sigma=1$,
$n=1000$) for the three values of $\lambda$ ($\nu=\log_{10}\lambda$).
The boldest line in Fig.~\ref{fig:17} corresponds to $\lambda=1000000$
($\nu=6$), and the thinner one to $\lambda=1000$ ($\nu=3$). The
solid horizontal 'significance levels' define the double error
band $\pm 2 Err[\bar{r}(t)]$ in case of $\nu=6$, and similar
dotted lines, for $\nu=3$ and $\nu=9$. In contrast to significance
levels of correlation coefficients, we have a smooth variable
$\bar{r}(t)$, which may for some time dwell outside the band
defined by the statistical error. Nevertheless, the relations
(\ref{hp_err_aver}) indeed characterize the behavior of typical
fluctuations of a smoothed variable for random data.

\begin{figure}[H]
\centering
\includegraphics{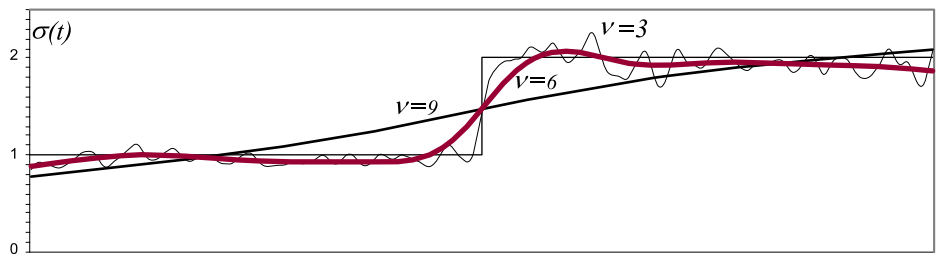}
\caption{Smoothed
volatility of process with step-constant $\sigma(t)$}\label{fig:19}
\end{figure}

However, in the non-stationary situation, which is a matter of our
main interest, we should keep the smoothing factor on the balance.
For example, if we model the process $\sigma(t)=1+0.5\cdot \sin
(2\pi t/T)$, where $T$ is the total duration of the simulated data
series, we get the following behaviors of volatility smoothing
(where volatility is measured by way of modified price range).
\begin{figure}[H]
\centering
\includegraphics{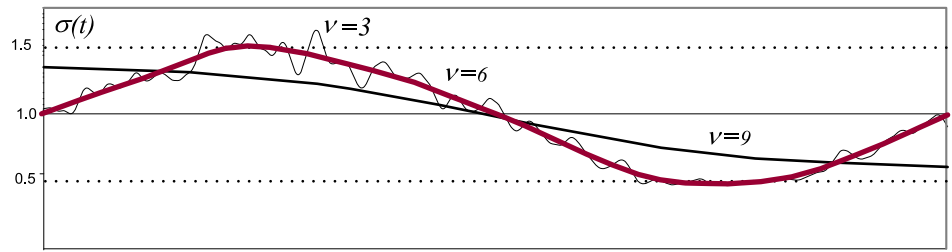}
\caption{Smoothed volatility of random 
walk with $\sigma(t)=1+0.5\cdot\sin (2\pi t/T)$}\label{fig:18}
\end{figure}
One can see from Fig.~\ref{fig:18} that in this case the optimal value 
is $\nu=6$, as $\nu=3$ follows too closely the noisy fluctuations around 
the true 
volatility, while $\nu=9$ simply does not 'catch' the periodic nature
of $\sigma(t)$. However, the
situation deteriorates dramatically, if volatility suffers a shock jump. 
Thus, let us consider the process, where for half of $n=1000$ 'trading days' 
the volatility is $\sigma=1\%$, and for the second half $\sigma=2\%$. For this 
model, HP-smoothing with different $\lambda$ gives the results plotted
in Fig.~\ref{fig:19}.
We see that in this case the choice of $\nu=6$ blurs the step significantly.
On the other hand, smoothing with $\nu=3$ approximates the jump in volatility
much better, but produces noisy and spurious fluctuations for constant $\sigma$.

\section{Autocorrelation of normalized volatility}

Let us now use the HP-filter to separate the smooth non-stationary
part of volatility and filter it out from the data. We will focus
on the higher-frequency component of volatility that remain after 
such filter is applied, as well as on the corresponding 
autocorrelation coefficients.

\begin{figure}[H]
\centering
\includegraphics{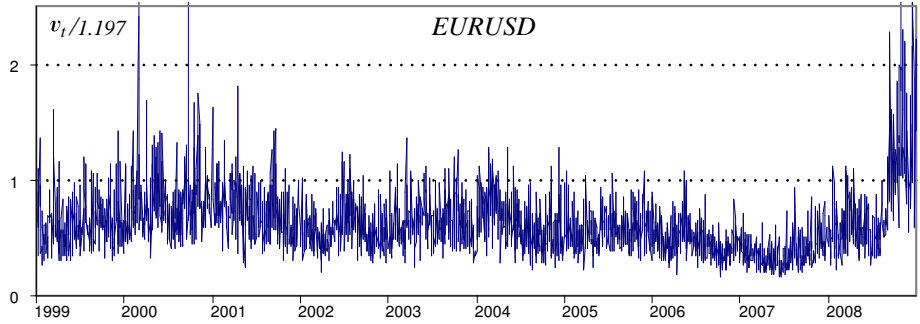}
\caption{Volatility of
EUR/USD measured by price range}\label{fig:20}
\end{figure}

Let us consider the daily modified price range
$v_t=a_t-|r_t|/2$ for EUR/USD exchange rate for the period from
1999 to 2008. Using this empirical data, we now estimate daily volatility
$\sigma_t=v_t\sqrt{2\pi}/3$ and plot it in Fig.~\ref{fig:20}.

\begin{figure}[ht]
\centering
\includegraphics{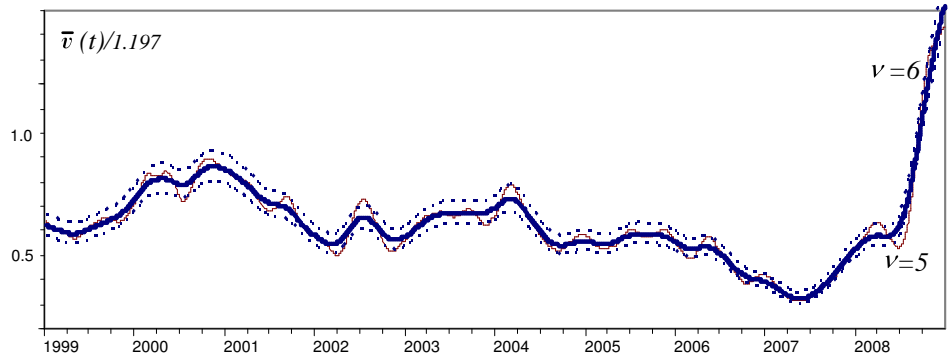}
\caption{Smoothing the
volatility with HP-filter}\label{fig:21}
\end{figure}

We extract the non-stationarity from the price process using HP-filter. 
The bold line at the chart below represents the volatility smoothed
with $\lambda=1000000$ ($\nu=6$). The double error band, according
with the equation (\ref{hp_err_aver}), for the value of volatility of 0.5 
(the average for years 2004-2007), will have the width of $\pm0.026$. In
fact, it is only slightly wider than the width of the line.
Therefore, the curves in the graph of non-stationary volatility $\sigma(t)$
for $\nu=6$ can be regarded as statistically significant (see Fig.~\ref{fig:21})

The picture changes when smoothing is performed with $\nu=5$
parameter. Let us take the graph for $\sigma(t)$ smoothed with
$\nu=6$, and plot the double error band $(1 \pm
0.036)\cdot\sigma(t)$ around it (marked by dotted lines), which
corresponds to the significance levels for $\nu=5$. As can be
observed from the chart, the $\nu=5$ smoothed volatility (thin
line) is more curvy than the one for $\nu=6$; however, all the
bends of the graph lie within the double-error corridor, and thus
one could assume they are not statistically significant. On the
other hand, the $\nu=5$ smoothed volatility models noticeably
better the behavior of the empirical data around the shock point
in fall of 2008.

As can be seen from the previous section, the HP-filter keeps the
curvature of the whole curve as low and as constant, as possible.
Therefore, it gives good results for relatively quiet intervals,
while producing larger distortion when the process goes through abrupt changes.

\begin{figure}[ht]
\centering
\includegraphics{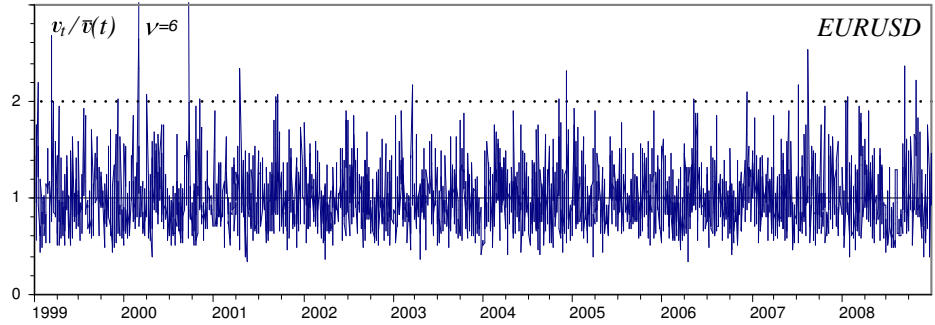}
\caption{Price ranges after normalization}\label{fig:22}
\end{figure}

\begin{figure}[hb]
\centering
\includegraphics{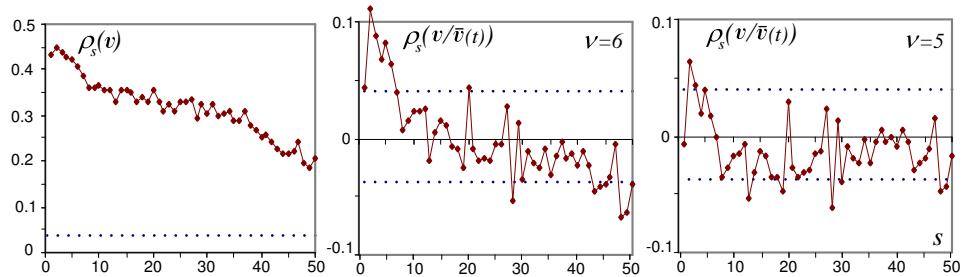}
\caption{Autocorrelations before and after normalization}\label{fig:23}
\end{figure}

Now we proceed to eliminate the smooth trend $\sigma(t)$ from the data.
We do this not by subtracting it, as is common practice in the
time series processing, but rather divide by it:
\begin{equation}\label{amplitude_normalization}
           \sigma_t \to \frac{\sigma_t}{\sigma(t)}.
\end{equation}
The meaning of this procedure is clear; it ensures that the volatility is 
normalized
for the entire data series. As a result of this procedure, the volatilities 
adjust 
not only their average, equal to 1, but also their variance, as can be
readily seen from Fig.~\ref{fig:22}.
Let us now compare the autocorrelation coefficients before the
normalization procedure~(\ref{amplitude_normalization}) is applied (Fig.~\ref{fig:23},
left), and after it is applied (Fig.~\ref{fig:23}, center and right).
As can be seen, the normalization reduces autocorrelations by nearly 10-fold. 
The same is true for the first correlation coefficient, which 
for the price range differences is equal to -0.50. Thus, its origin
is indeed related to the effect of overlap discussed above.

We note that during the normalization procedure we divide all daily
amplitudes by the smoothed variable $\sigma(t)$. However, when
we calculate it, we use a set of values $\sigma_t$ at and around the
current time $t$. As a result, the neighboring values of
$\sigma(t)$ could appear significantly correlated. This may lead to
small autocorrelation present after normalization; nevertheless,
the value of $\rho_s(v/v(t))$ is very small.

Thus, both simple transition to first differences of the data series, 
and removal of the smooth component of volatility by means of HP-filter, 
make correlation coefficients of the adjusted process statistically insignificant. 
This fact, combined with the discussed above simple explanation for the origin 
of autocorrelation under non-stationarity, raises doubts about the stochastic 
nature of volatility. 
However, one still needs to explore in more depth the noisy component of the
volatility. We will return to this issue in the last section of the paper.

\section{Back to normal distribution}

As was already mentioned in the Introduction, there is a large
body of research that study the probability distribution
of logarithmic returns. The fact of its being non-Gaussian has become
generally accepted (see, for example \cite{Jondeau:2007}, \cite{Barndorff:2001}).
However, when we speak of the density of probability as function
of single variable $P(r)$, we obviously assume the stationarity of
random numbers $r$, as we do not involve time dependency. To
obtain sufficiently reliable statistical results when inferring
$P(r)$, one chooses the widest possible interval containing a
large amount of data points $n$.

However, under non-stationarity such approach significantly distorts
the 'true' type of distribution. If statistical parameters
depend on time, the density of distribution will not be stationary either $P(r,t)$. 
Let us assume that the
non-stationarity is {\it parametric} and concentrated only in the
volatility $\sigma(t)$. Suppose also that $P(r,t)=P(r,\sigma(t))$
is governed by the Gaussian distribution ($r_t=\sigma(t)\cdot \varepsilon_t$):
\begin{equation}
       P(r,t) = \frac{1}{\sigma(t)\sqrt{2\pi}} ~e^{-\frac{1}{2}~ r^2/\sigma^2(t)}.
\end{equation}
Second and forth moments are equal to, respectively: 
$\overline{r^2} = \left<\sigma^2(t)\right>,~\overline{r^4} = 3\left<\sigma^4(t)\right>$,
and in general case, despite the Gaussian distribution, its
'aggregated' kurtosis, estimated without taking into account the
non-stationarity, becomes different from zero:
\begin{equation}
      ex = 3\cdot \left[ \frac{\left<\sigma^4(t)\right>}{\left<\sigma^2(t)\right>^2}- 1 \right].
\end{equation}
In our toy model of a 20-year walk with shock volatility doubling,
the kurtosis of data equals to 27/25 = 1.08. In a more general case,
the non-Gaussian nature may be affected by other types of non-stationarity, 
for example, the drift of returns: $r_t=\mu(t)+\sigma(t)\cdot\varepsilon_t$.

Let us see what happens with the empirical data after eliminating of
the non-stationarity. In order to do this we divide all $r_t$ 
by the value of volatility at a given moment of time. We
obtain its current value by smoothing daily modified amplitudes of range 
$\sigma_t=(a_t-|r_t|/2)\sqrt{2\pi}/3$ using the HP-filter.
Thus, we apply the following transformation to initial logarithmic returns:
\begin{equation}\label{transform_normalization}
     r_t \to r'_t=  \frac{r_t}{\sigma(t)}.
\end{equation}
Such normalization makes random numbers $r'_t$, modulated by
$\sigma(t)$ function, stationary.

Table~\ref{tbl:4} contains statistical parameters of S\&P500
index logarithmic returns for the period 1990-2008. The total
number of trading days is equal to $n=4791$, the share of positive
returns is 52.8\% for all cases.
\begin{table}[H]
\centering
\small
\caption{Statistical parameters of S\&P500 index logarithmic returns for three different degrees of smoothness parameter $\nu$}
\label{tbl:4}
\begin{tabular}{rrrrrrrrrrrrrrrrrrr}
              &     $aver$    &   $sigma$ &   $asym$ &    $excess$  &  $p_1$  \\
\hline
$r$          &   0.020    & 1.137   &  -0.23 &     10.18  &  78.9   \\
$\nu=6$      &   0.051   &  1.199 &  -0.15 &   1.19  &  71.1 \\
$\nu=5$      &   0.055     &  1.187 &  -0.12 &      0.82  &  70.4 \\
$\nu=4$      &   0.059    &  1.177 &  -0.08 &   0.51  &  69.5 \\
\end{tabular}
\end{table}
The first line presents the statistics before the
transformation of normalization (\ref{transform_normalization}).
The other lines contain statistic parameters after transformation, where
smoothing with differing parameter $\nu=\log_{10}\lambda$ is used.

Special attention should be paid to the columns $excess$ and $p_1$. 
We see that
smoothing reduces drastically the values of these parameters.
This is true even for a sufficiently smooth function $\sigma(t)$,
corresponding to $\nu=6$. In Fig.~\ref{fig:24} it is represented by the bold
line:
\begin{figure}[h]
\centering
\includegraphics{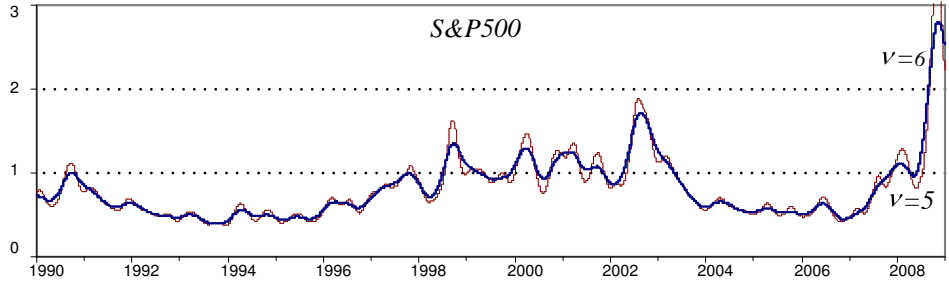}
\caption{Smooth
volatility of S\&P500 for different parameters $\nu$}\label{fig:24}
\end{figure}

\begin{figure}[H]
\centering
\includegraphics{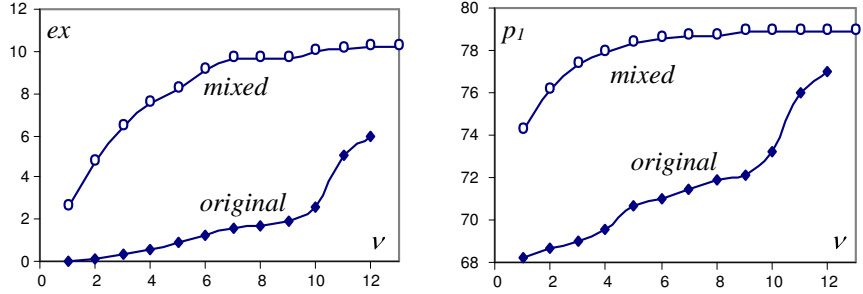}
\caption{Kurtosis $ex$
and probability $p_1$ for different $\nu$, 1990-2008}\label{fig:25}
\end{figure}

The smaller parameter $\nu$ is, the more intensive the flections of
volatility $\sigma(t)$ are, because the fluctuations of returns
start affecting the average. Obviously, in this case a decrease in
kurtosis takes place, even for stationary non-Gaussian random
process. To control this effect, we perform the following
simulation experiment. We randomly mix the initial pairs of daily 
returns and
volatility $\{r_t, \sigma_t\}$  in order to eliminate
non-stationarity. After that, we apply smoothing with HP-filter,
and normalization (\ref{transform_normalization}) both to initial
data (original), and to mixed ones (mixed). The charts in Fig.~\ref{fig:25}
present the dependence of the kurtosis (left) and the probability
of the fact that returns fall within one sigma $p_1$ (right) as
functions of the smoothing parameter $\nu=\log_{10}\lambda$.
It can be easily noticed that to the right of $\nu\sim6$ the
kurtosis and probability $p_1$ for mixed data decrease 
insignificantly. At the same time, statistical parameters 
characterizing the non-Gaussian property of initial data decrease rapidly.

Thus, as the criterion for the optimal meaning of $\nu$, one may choose 
the point where the difference between the statistics of mixed and
initial data reaches its maximum.

Another argument for importance of non-stationarity contribution
into the non-Gaussian property of distribution is the
break out of 2008 financial crisis. As can be seen from Table~\ref{tbl:4},
the kurtosis over the period 1990-2008 is equal to $ex=10.2$.
However, it is enough to eliminate just one volatile year of 2008, in
order to make the kurtosis decrease threefold to $ex=3.8$. 
The number of trading days for this calculation is reduced in this case by 
only 5\% to $n=$4528.

\begin{figure}[hb]
\centering
\includegraphics{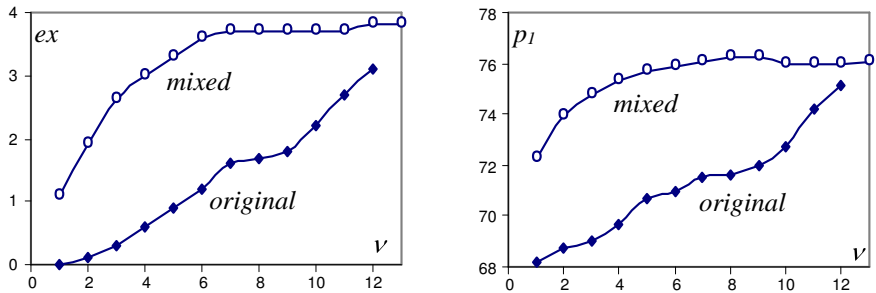}
\caption{Kurtosis $ex$
and probability $p_1$ for different $\nu$, S\&P500, 1990-2007}\label{fig:26}
\end{figure}
The charts in Fig.~\ref{fig:26} depict the dependency of kurtosis and probability $p_1$
on the smoothing parameter $\nu$ for mixed and initial data of
S\&P500 index daily returns for the period 1990-2007.
One can notice that, although the initial value of kurtosis is
relatively small, it nevertheless decreases statistically
significantly as a result of elimination of non-stationarity from the
data. For normalized data, the value of kurtosis $ex=1$ can be considered
as significant, which is four times smaller than for
initial data.

\begin{figure}[H]
\centering
\includegraphics{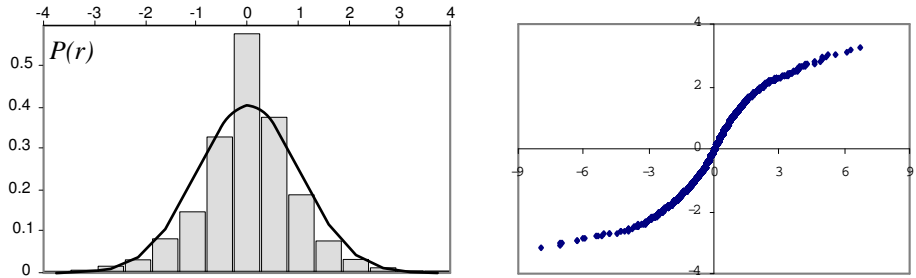}
\caption{Distribution
of S\&P500 returns 1990-2008}\label{fig:27}
\end{figure}

\begin{figure}[ht]
\centering
\includegraphics{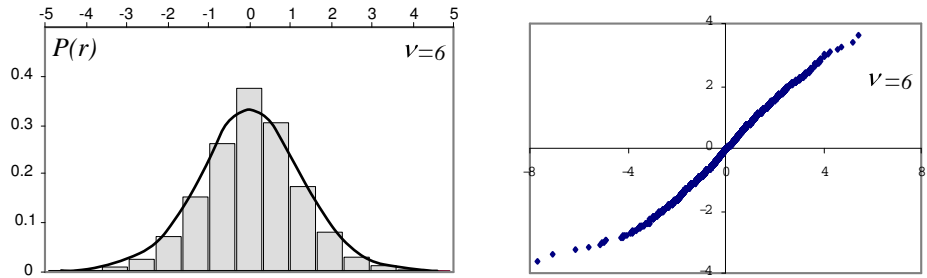}
\caption{Distribution
of S\&P500 returns for 1990-2008 after normalization }\label{fig:28}
\end{figure}

Let us plot (see Fig.~\ref{fig:27}) the histograms of probability density
distribution and a graph of normal probability (in a way similar
to \cite{Fama:1965}),
formally based on the initial non-stationary data,
as well as the same quantities after the normalization procedure
(\ref{transform_normalization}) is applied to the data (Fig.~\ref{fig:28}).
The unmarked line in the charts corresponds to the Gaussian
distribution. A graph of normal probability represents dependency
$y=f(r)$ of relation $F_{{\bf N}}(y)=F(r)$, where $F_{{\bf N}}(y)$
is an integral normal distribution, and $F(r)$ is empirical
integral distribution for returns. If the empirical distribution
of $F(r)$ is Gaussian, this graph should be a straight line.
We see that after normalization the density of probability
becomes much more close to normal. Deviations from the straight line 
are particularly evident for the excessively large negative returns because
of the rare negative shock impacts to the market.

Let us consider, for comparison, the probability distribution of
currency market daily returns using the EUR/USD rate for
the period 1999-2008 as sample. Basic statistical parameters before
normalization (first line) and after smoothing with different
parameters $\nu$ are given in Table~\ref{tbl:5}.
We see that the initial data has relatively small kurtosis, but
after smoothing it decreases even further. The mean value of
volatility after normalization is close to one. This confirms that
$\sigma=v\sqrt{2\pi}/3$ is a good unbiased estimation of the
daily volatility of rate returns.
\begin{table}[H]
\centering
\small
\caption{Statistical parameters of EURUSD rate logarithmic returns for three different degrees of smoothness parameter $\nu$}
\label{tbl:5}
\begin{tabular}{rrrrrrrrrrrrrrrrrrr}
              &     $aver$    &   $sigma$ &   $asym$ &    $excess$  &  $p_1$  \\
\hline
$r$          &   0.008   &  0.652   &    0.05   &     1.3    &   72.7 \\
$\nu=10$     &   0.017   &  1.022   &    0.03   &     0.8    &   71.5 \\
$\nu=~6$     &   0.022   &  0.995   &    0.00   &     0.1    &   69.3 \\
$\nu=~4$     &   0.022   &  0.993   &    0.01   &     0.1    &   69.0 \\
\end{tabular}
\end{table}

\begin{figure}[htb]
\centering
\includegraphics{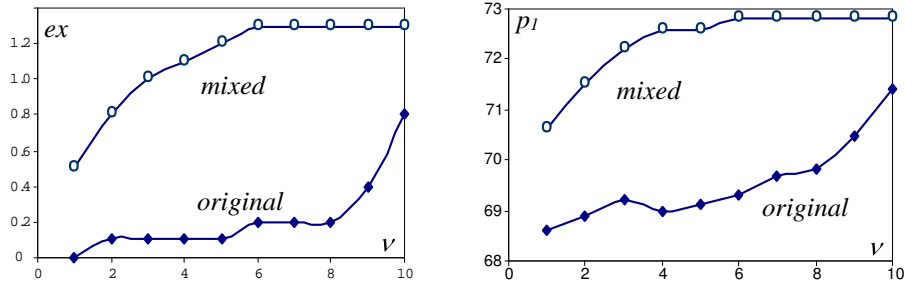}
\caption{Kurtosis $ex$
and probability $p_1$ for different $\nu$, EURUSD, 1999-2008}\label{fig:29}
\end{figure}

\begin{figure}[htb]
\centering
\includegraphics{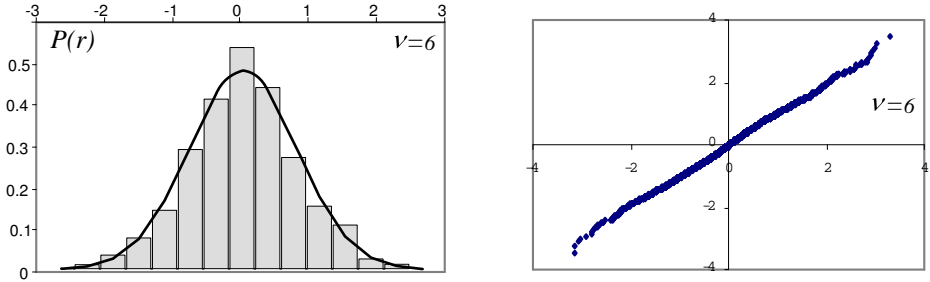}
\caption{Distribution
of returns EURUSD, 1999-2008, after normalization }\label{fig:30}
\end{figure}

Testing statistical significance of the decrease in kurtosis
and the probability $p_1$ shows practically zero kurtosis of
normalized returns (Fig.~\ref{fig:29}).
The corresponding histogram and normal probability graph are plotted in
Fig.~\ref{fig:30}.
As a result we receive a virtually canonical normal distribution
with deviations that are rather typical for a relatively small
sample ($n=2495$).

We shall not conduct a more detailed statistical analysis of
distribution form, limiting the argumentation to these illustrative 
examples. We infer (see Conclusion) that the observed data is composed 
of the mixture of
normally distributed fluctuations of the market, modulated with
non-stationary volatility, and rare shock impacts. Therefore, even
after the elimination of non-stationarity there may remain shock
outliers, which make the total distribution weakly non-Gaussian.

\section{Quasi-stationarity of volatility}

Vanishing autocorrelation coefficients between consecutive values of
volatility, generally speaking, do not exclude the possibility of
its stochastic description. In particular, we can write down the
following simple discrete process:
\begin{equation}
    r_t = \sigma_t \cdot \nu_t,~~~~~~~\sigma_t = \sigma\cdot(1+\beta\cdot\mu_t),
\end{equation}
where $\nu_i$ and $\mu_i$ are independent random variables, while
$\sigma$, $\beta$ are constants. However, within this model the
interpretation of volatility $\sigma_t$ as a random variable
becomes rather superfluous. In fact, we come back to the usual
stationary model $r_t=\sigma \varepsilon_t$, where
$\varepsilon_t=\nu_t + \beta \cdot \mu_t\nu_t$. In particular, if
$\nu_i$ and $\mu_i$ are normally distributed, the distribution for
$\varepsilon_i$ would no more be normal with kurtosis equal to
$6\beta^2(2+\beta^2)/(1+\beta^2)^2$. Nevertheless, the question of
{\it local} stationarity of 'true' volatility remains open.

Let us conduct several statistical estimations. First, we
consider a modified amplitude of range. The spread of its
values under constant volatility $\sigma$ occurs due to finite
width of distribution density $P(v)$. One can obtain its analytical
form from the equation (\ref{appendix_P_a_r}) of \ref{appxA}, and
present it as the following infinite series:
\begin{equation*}
 P(v)=(32 v^4-9){\bf N}(2v)
     +\sum^\infty_{k=2}\left\{ \frac{4(2k-1)^2}{k^2(k-1)^2}{\bf N}_1
     -\frac{8k^2\bigl(1+k^2-4(k^4-k^2)v^2\bigr)}{(k^2-1)^2}{\bf N}_2\right\},
\end{equation*}
where ${\bf N}_1={\bf N}(2(2k-1)v)$, ${\bf N}_2={\bf N}(2kv)$ are
non-normalized Gaussian functions (see \ref{appxA}). Below we list the
integral probabilities of the fact that variable
$\sigma=v\sqrt{2\pi}/3$ falls within the interval $[0..\sigma_0]$
(the first line contains values of $\sigma_0$, the second, corresponding
probabilities measured in percent points):
\begin{center}
\small
\begin{tabular}{rrrrrrrrrrrrrrrrrrr}
\hline
0.5  &  0.6  &  0.7  &  0.8  &  0.9  & 1.0      &  1.1    & 1.2  &  1.3   & 1.4  &  1.5  & 1.6\\
0.6  &   3.3 &   10.4 &  22.5 &  37.8  & 53.7 &  68.1  & 79.5 &  87.7  & 93.1  & 96.4  & 98.2\\
\hline
\end{tabular}
\end{center}
The modified price range $v\sqrt{2\pi}/3$ should remain within the
interval $[0~..~1.5]$ about 96.4\% of days; it very rarely drops below 0.5.

\begin{figure}[H]
\centering
\includegraphics{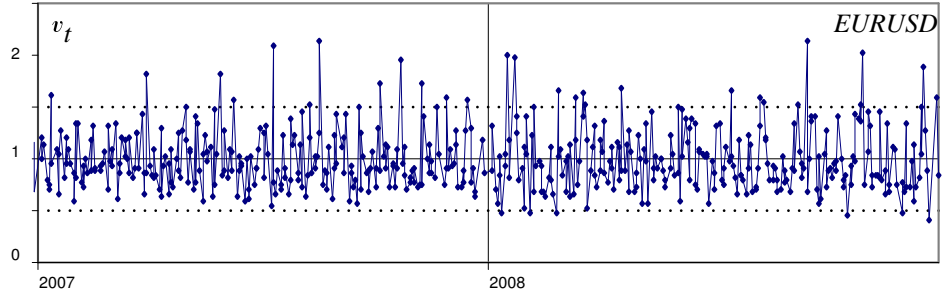}
\caption{Stationary
price range of EURUSD in 2007-2008}\label{fig:31}
\end{figure}

If we eliminate (\ref{transform_normalization}) by smoothing procedure with
$\nu=4$ the non-stationarity in daily modified ranges for
EURUSD in 2007-2008 years, the residual series has dynamics as shown in
Fig.~\ref{fig:31}.
In case of Brownian random walk, dotted lines correspond to the
probability 96\% of staying within the interval 
$0.5 < v\sqrt{2\pi}/3 < 1.5$. We see that, except for rather rare
outliers, most of daily volatilities estimated by modified
amplitude of probability, fell into the dotted corridor. The number
of outliers is slightly higher than expected 4\% (as there is
250 trading days in a year, 250*4\%=10). This small excess of
extremal values may be interpreted (especially in 2008, a crisis year) 
as occasional shock impacts to the market, not
related to its 'typical' intrinsic dynamics.

\begin{figure}[ht]
\centering
\includegraphics{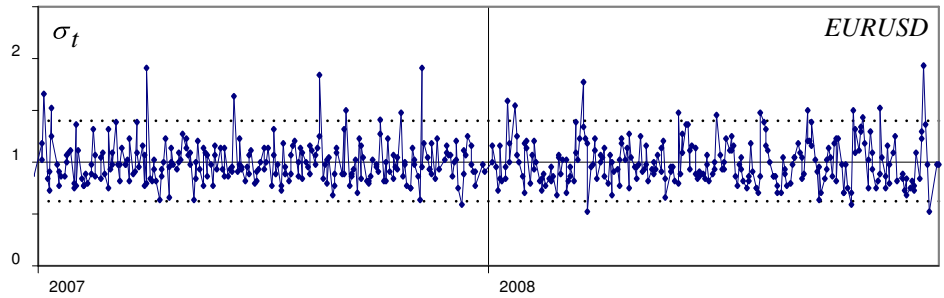}
\caption{Intraday
volatility of EURUSD in 2007-2008}\label{fig:32}
\end{figure}

As we have discussed above, the 'daily' volatility can be
estimated not only by means of modified amplitudes of range,
but also by calculating its value on the base on intraday lags, i.e.
15-minute ticks. In Fig.~\ref{fig:32}, the dynamics of volatility is presented, 
after the elimination of non-stationarity, obtained by the latter method
for the period of 2007-2008 for EUR/USD exchange rate.
In this case the spread of values is related to the finiteness of
the sample that is used for volatility calculations. In order to
determine the significance level, one has to know the
corresponding distribution of probability. As we know, the error of
stationary volatility calculation is determined by fourth moments
and, in case of large kurtosis, it will be quite large.

The intraday 15-minute data distribution has significant kurtosis.
Straightforward computation of kurtosis for EURUSD during 2004-2008
yields the value of 20, which is due to the long-term non-stationarity,
the substantial cyclic effects in intraday activity, as well as
several other specific reasons (see \cite{Biais:2005},
\cite{Madhavan:2000} for a detailed discussion), 
into which we will not delve here.

In order to obtain the significance levels, we will conduct the
following simple experiment with the data. Let us calculate the
logarithmic return basing on 15-minute lags of EUR/USD rate. Then,
to preserve the intraday periodicity, we shall mix data points with the
same intraday time. In other words, we randomly shuffle
all lags at 00:00, then apart from them we mix lags of 00:15, etc.
For these synthetic data, that are free of any memory effects except the
intraday cycles, we calculate the meaning of intraday volatility.
Then we normalize the series so that the mean is equal to
1, and plot the corresponding distribution of probability. It turns
out that about 96\% of data stays within the $1\pm 0.4$ corridor.
It is these levels, which characterize the 'typical' range of
volatility due to the finiteness of data, that are marked at the
above chart with dashed lines. We see that the data fits into the corridor
quite well.

We stress that the computations performed above are rather a
qualitative estimation than a strict statistical analysis; such analysis
might not be appropriate at all without constructing a complete model of
non-stationarity of data at different time scales. However, an
assumption about the local constancy of volatility of daily lag
returns appears rather plausible. In other words, the daily
volatility of the market most probably is described by a smooth,
rather slowly varying function of time. At any moment, its value
can be considered locally constant, and it determines the
stochastic dynamics of price returns for a financial instrument.

\section{Conclusion}

Let us reiterate the main inference that we argued for in this paper:
\begin{quote}
Volatility and other statistical parameters should be regarded as
gradually changing functions of time. They determine locally
quasi-stationary stochastic dynamics of prices for financial
instruments. There are {\em rare} and {\em irregular}
shock impacts that influence the markets, resulting in shifts in
daily returns, and as they accumulate, affect the value of long-term 
volatility.
\end{quote}

The situation resembles the deformation of a plastic material
after a series of impacts, and the gradual restoration of form 
after external influence is terminated. The study of properties of 
such resilience of
volatility are of great importance, especially for forecasting 
the time of its reversal to the long-term typical levels.

Therefore, the stochastic nature of markets is determined by the
following two components: 1) intrinsically Gaussian-distributed
daily returns with slowly changing volatility; and 2) rarely
occurring shock impacts. These shocks are assumed to be
essentially unpredictable, but their impact on the volatility as
well as its subsequent evolution should be the subject of
research.

Actually, shocks are quite inconspicuous; in reality it is quite difficult 
to separate the 'unnatural' behavior of the market as a result of shocks
from 'normal' volatility. Financial analysts and economic commentators
never fail to find the piece of news to account for all price spikes and 
crashes. On the other hand, such events as Lehman's bankruptcy can 
hardly be considered everyday news.

Volatility can also gradually increase as a result of relatively
insignificant negative news background, provided that such background 
lasts for long enough. Thus, a gradual increase in volatility since the
beginning of 2007 was a result of precisely such 'soft' 
pressure on the markets from the real estate sector. Since the
autumn of 2008, this growth has been explosive and unprecedented
for the modern history of financial markets. As we know, it
originated from financial sector, and triggered an avalanche-like
effect of confidence crisis and widespread panic. All this, 
eventually, delivered a blow to the real sector of economy.

Finally, an increase in volatility usually accompanies 
'unmotivated' booms in the market, when a financial bubble starts
to inflate. High volatility also persists in the period of its 
collapse. When market goes into a 'quiet' phase
of growth, volatility usually slowly decreases.

Peaks typically observed in the charts of non-stationary volatility 
bring up
the analogy with resonance phenomena in physics. Such connection
implies the existence of certain equations describing the system dynamics.
There is no doubt that a relaxation mechanism exists, ensuring
that a decay of system excitations happens after a certain period,
determined by the life time of the resonance.

When one speaks about a gradual course of change in volatility, one should
keep in mind that it refers to the 'typical' long-term market situations. 
Sometimes, however, jump-like changes in statistical parameters occur, 
which determine the stochastic dynamics of the price process. 
It seems plausible that such a qualitative shift in market behavior happened 
in September 2008. In contrast, the exit from this instability, and return to
equilibrium, is likely to be quite gradual and prolonged.

We infer that the non-Gaussian nature of markets stems from two
origins. First, it is the artifact of uncritical postulation of
stationarity under conditions when it doesn't really exist. This
component can be removed, at least in theory.
After the data is transformed into a stationary form, the non-Gaussian 
features reduce significantly.
However, the rare shock impacts, which are the second origin, 
even when combined with stationary
Gaussian returns, still render the distribution weakly non-Gaussian. 
This is particularly evident in the case of stock market, which has the
after-hours periods when negative or positive news accumulate.
When the markets open, a possibility appears of a 'single
emission' of accumulated emotions. Around-the-clock foreign exchange
markets can respond to the development of such shocks in more 
'subdued' way.

Autocorrelation coefficients in various volatility measures also
arise due to the non-stationarity of the data and disappear after
it is eliminated. In this sense, they are indeed the evidence of
long-term memory, but do not have anything to do with the
short-term stochastic properties of volatility, which are assumed
in corresponding autoregressive models. Therefore, further
research should focus on forecasting the smooth dynamics of
volatility, rather the stochastic theories of volatility behavior.

\section*{Acknowledgement}

I am grateful to
Alexander Zaslavsky, Igor Chavychalov, Andrej Tishchenko,
Oleg Orlyansky, Leonid Savtchenko, Alexander Ferludin and
Anna Gorbatova for many useful comments.
Any remaining errors are my own.

\newpage

\appendix

\setcounter{section}{0}
\renewcommand{\thesection}{Appendix \Alph{section}}
\renewcommand{\theequation}{\Alph{section}\arabic{equation}}
\setcounter{equation}{0}

\section{Brownian walk}\label{appxA}

In this appendix we provide the basic expressions for Brownian motion 
described by the
stochastic equation $dx = \mu dt + \sigma \delta W$. Let us first
consider the case of driftless process ($\mu=0$). Without any loss of
generality, we may assume that at the initial moment of time
$x(0)=0$. The maximal and minimal values of $x$ for the period
$0\leqslant t \leqslant T$ are equal to $H$ and $L$, respectively, 
and $r=x(T)$.
The height $h=H$ of ascent and the depth $l=-L$ of descent are always
positive, and $-l \leqslant r \leqslant h$. The amplitude of range
is equal to $a=h+l$. Below we consider the case of unit volatility
$\sigma=1$ and unit time interval $T=1$. To restore the original
notation, it
is necessary to substitute $r\to r/\sigma\sqrt{T}$
for the dimensionfull variables $r$, $h$, $l$, $a$. The same should
be done in the differentials $dr$, etc. in the integrals containing the
probability densities. In order to make the formulae more concise, we use this
notation for the normal
distribution function: ${\bf N}(x)=e^{-x^2/2}/\sqrt{2\pi}$.

$\bullet$ We start with the relation for probability that $x$ does
not rise above $h$ and does not fall below $-l$, when the closing
return is $r$:
\begin{equation}\label{Feller_prob_HL}
p(-l<L, H<h, r) =
\sum^{\infty}_{k=-\infty}
\left\{ {\bf N}\bigl(r+2ka \bigr)-{\bf N}\bigl(r+2l+2ka\bigr) \right\}.
\end{equation}
This formula was first received by \cite{Feller:1951}. We
also note an exclusively useful reference book by
\cite{Borodin:2000}. Distributions for other variables are derived 
from the
probability (\ref{Feller_prob_HL}). For the return, height
and depth we have:
\begin{equation}
P(r) = {\bf N}(r),~~~~~P(h)=2{\bf N}(h),~~~~~~P(l)=2{\bf N}(l).
\end{equation}
The density of probability for the range $a$ is expressed in the
form of an infinite series over Gaussian basis:
\begin{equation}
P(a)=8\sum^\infty_{k=1}(-1)^{k+1}\cdot k^2\cdot {\bf N}(ka).
\end{equation}
This series converges rather quickly for all $a\neq 0$. A characteristic
property of Feller's distribution $P(a)$ is an extremely rapid
decline in the density of probability for large values of $a$. 
Here is a sample of values of integral probabilities
$F(a)=p(H-L<a)$:
\begin{center}
\small
\begin{tabular}{rrrrrrrrrrrrrrrrrrr}
$a$     & 0.750 & 1.000  &  1.500  &  2.000  & 2.500  & 3.000 & 3.500 & 4.000 \\
$F(a)$  & 0.002 & 0.063  &  0.487  &  0.819  & 0.950  & 0.989 & 0.998 & 1.000 \\
\end{tabular}
\end{center}
The $a$ parameter is smaller than 0.75 ($\sigma=1$) only in 2 cases
out of 1000. The mean value is $\bar{a}=1.5958$, the variance is
$\sigma_a = 0.29798\cdot \bar{a}$. 
The one sigma interval ($\bar{a}\pm \sigma_a$ = [1.120~..~2.071]) contains
71.6\% of all $a$ values, while the double sigma interval
($\bar{a}\pm 2\sigma_a$ = [0.645~..~2.547]) contains 95.6\%; 
and data points outside of the latter interval should, in reality, 
{\it occur only for $a$ above the mean}.

The joint densities of probability for height ($r \leqslant h$),
depth ($-l \leqslant r$) and range ($|r| \leqslant a$) have
the following form:
\begin{equation}
P(h, r) = 2(2h-r)\cdot {\bf N}(2h-r),~~~~~P(l,r)=2(2l+r)\cdot{\bf N}(2l+r).
\end{equation}
\begin{equation}\label{appendix_P_a_r}
P(a, r) =
4\sum^{\infty}_{k=-\infty}k\cdot
\Bigl\{-|r|-k(2k+3)a+k\cdot\bigl(a-|r|\bigr)\bigl(2ka+|r|\bigr)^2\Bigr\}\cdot {\bf N}\bigl(|r|+2ka\bigr).
\end{equation}
Note also that $P(a, -r)=P(a, r)$, and $P(a, |r|)=2P(a,r)$.

$\bullet$ Let us provide a table of mean values for different
variables (where $v=a-|r|/2$):
$$
\begin{array}{llllll}
\overline{r} = 0,&
\overline{r^2}= 1,&
\overline{r^3}= 0,&
\overline{r^4} = 3,\\
\\
\displaystyle
\bar{h} = \sqrt{\frac{2}{\pi}},&
\overline{h^2} =1,&
\displaystyle
\overline{h^3}=\sqrt{\frac{8}{\pi}},&
\overline{h^4} = 3,\\
\\
\displaystyle
\overline{a} = \sqrt{\frac{8}{\pi}},~~~ &
\displaystyle
\overline{a^2} = 4\ln 2,~~~~  &
\displaystyle
\overline{a^3} = \frac{(2\pi)^{3/2}}{3},~~~~ &
\displaystyle
  \overline{a^4} = 9 \cdot \zeta[3], \\
\\
\displaystyle
\overline{v} = \frac{3}{\sqrt{2\pi}},~~~ &
\displaystyle
\overline{v^2} = 4\ln 2 - \frac{5}{4},~~~~  &
\displaystyle
\overline{v^3} = \frac{21+\pi^2}{6\sqrt{2\pi}},~~~~ &
\displaystyle
  \overline{v^4} = 6\ln 2 -\frac{27}{16}+ \frac{3}{8}\cdot \zeta[3], \\
\end{array}
$$
where $\zeta[n]= \sum^\infty_{k=1} k^{-n}$ is a Rieman $\zeta$-function.
The mean values for $l$ and $|r|$ are the same as for $h$. The means of certain
cross-products are given below:
$$
\begin{array}{cccc}
\displaystyle
\overline{h\, r} =\frac{1}{2},~~&
\displaystyle
\overline{h\, r^2} =\frac{4}{3}\sqrt{\frac{2}{\pi}},~~&
\displaystyle
\overline{h\, r^3} =\frac{3}{2},~~ &
\displaystyle
\overline{h\, r^4} =\frac{24}{5}\sqrt{\frac{2}{\pi}},~~\\
\\
\displaystyle
\overline{l\, r^n} =(-1)^n\cdot\overline{h\, r^n},~~ &
\displaystyle
\overline{a\, r^{2n+1}} = 0,~~&
\displaystyle
\overline{a\, r^{2n}} = 2 \cdot \overline{h\, r^{2n}},~~&
\displaystyle
\overline{a \,|r|} =\frac{3}{2}.~~\\
\end{array}
$$
Expressions for other mean values, as well as their generating function, 
can be found in \cite{Garman:1980}.

For a process with a non-zero drift $dx=\mu dt+\sigma \delta W$, we shall use
the above-determined driftless densities. In order to restore time $T$
and variance $\sigma$, we should additionally substitute the shift
as follows: $\mu\to\mu T/\sigma \sqrt{T}$. The density of probability for
returns is equal to:
$$
P_\mu(r)={\bf N}\bigl(r-\mu\bigr)=e^{\mu r - \mu^2/2} ~P(r).
$$
Expressions for joint probability densities \cite{Borodin:2000}:
$$
P_\mu(h, r)  = e^{\mu r - \mu^2/2}~P(h, r),~~~~~
P_\mu(l, r) = e^{\mu r - \mu^2/2}~P(l, r),
$$
$$
P_\mu(a, r) = e^{\mu r - \mu^2/2} ~P(a, r),~~~~~
P_\mu(h, l, r) = e^{\mu r - \mu^2/2}~ P(h, l, r).
$$
Thus, the densities corresponding to $\mu=0$ are always multiplied
by a factor $e^{\mu r - \mu^2/2}$. In the presence of drift we obtain:
$$
\overline{r}=\mu,~~~\overline{r^2}=1+\mu^2,~~~~\overline{r^3}=3\mu+\mu^3,~~~~~\overline{r^4}=3 + 6\mu^2+\mu^4.
$$
Exact expressions for mean values of other variables are rather cumbersome.
However, as for financial data the condition $\mu\ll\sigma=1$ holds, it is
acceptable to decompose a factor $e^{\mu r - \mu^2/2}$ into a series
and to use means for the case $\mu=0$. As a result we receive:
\begin{equation}\label{appendix_aver_h_r_mu}
\overline {h} =\sqrt{\frac{2}{\pi}} + \frac{\mu}{2} + \frac{\mu^2}{3\sqrt{2\pi}} - \frac{\mu^4}{60\sqrt{2\pi}}+..,
~~~~\overline{|r|} = \sqrt{\frac{2}{\pi}} + \frac{\mu^2}{\sqrt{2\pi}} - \frac{\mu^4}{12\sqrt{2\pi}} +..,
\end{equation}
\begin{equation}\label{appendix_aver_l_a_mu}
\overline {l} =\sqrt{\frac{2}{\pi}} - \frac{\mu}{2} + \frac{\mu^2}{3\sqrt{2\pi}} - \frac{\mu^4}{60\sqrt{2\pi}}+..,
~~~~\overline{a} = \sqrt{\frac{8}{\pi}} + \frac{2\mu^2}{3\sqrt{2\pi}} - \frac{\mu^4}{30\sqrt{2\pi}} +...
\end{equation}
The mean values of height and depth are linear in $\mu$, and only even powers
of $\mu$ are present in the tail of expansion. The means of lag range
and absolute returns contain only even powers of $\mu$. Note also the following
simple relations, available in closed form:

~~$\overline{h}-\overline{l}=\overline{r}=\mu$,
~~~$\overline{h^2}+\overline{l^2}=2+\mu^2$,
~~~$\overline{h\,r}=\overline{h^2}-1/2$, ~~ $\overline{l\,r}=1/2 -
\overline{l^2}$.

\section{Measures of volatility}\label{appxB}

The width of probability
distribution of a positive random variable $z>0$ can be 
characterized with a relative error
$\sigma_z/\bar{z}$, where $\sigma_z$ as usual denotes the
standard deviation $\sigma^2_z=\overline{(z-\bar{z})^2}$.

Note that the relative width of distributions for $z$ and $z^2$
are different, and thus actually there are different
criteria for optimality of volatility measurement. For example, in
order to calculate the stationary volatility one usually uses
averaging of either squared returns, or the squares of the
lag ranges \cite{Parkinson:1980}:
\begin{equation}\label{appendix_sigma_R_P}
 \sigma^2_R = \frac{1}{n-1}\sum^n_{t=1} (r_t-\bar{r})^2,~~~~~~~~~~~\sigma^2_P = \frac{1}{n}\sum^n_{t=1} \frac{a_t^2}{4\ln 2}.
\end{equation}
As in this paper we examine the non-stationary nature
of volatility and use the non-linear HP-filter for smoothing, it
is more convenient to average volatilities $\sigma$ proper,
rather than their squares; the latter, as we will see below, 
yield a
biased value of $\sigma$ for small $n$. Nevertheless, considering
the various measures of volatility, we will calculate the
relative width of both the value its square.

Let us recite some well-known volatility estimators. We shall use a
Parkinson measure (1980) \cite{Parkinson:1980} as a base; it
is equal to the amplitude of range $v_P=a$. Garman and Klass
(1980) \cite{Garman:1980}, working in the class of analytic functions of
$h$, $l$, $r$, proposed the following optimal combination, which
is a better measure than that of Parkinson:
\begin{equation}\label{appendix_sigma_RG}
   v^2_{GK} = 0.511 \cdot a^2 - 0.019 \bigl(r\cdot (h-l)+ 2 h\cdot l\bigr) - 0.383 \cdot  r^2.
\end{equation}
A simpler and drift-independent $\mu$ measure is suggested by
Rogers and Satchell (1991) \cite{Rogers:1991}:
\begin{equation}
   v^2_{RS} = h\cdot (h-r) + l\cdot (l+r).
\end{equation}

Let us show that the simplest {\it linear} modification of Parkinson's
measure
\begin{equation}\label{appendix_v_beta}
          v_\beta=a - \beta\cdot |r|
\end{equation}
where $\beta>0$ is a constant, leads to a narrower distribution
than the amplitude of range. If we use relative volatility
$\sigma_v$ as a criterion of narrowness, it is not difficult to 
find the optimal value of the
coefficient $\beta$ using the means from
the Appendix A:
\begin{equation}
    \frac{\overline{(a-\beta \cdot |r|)^2}}{\bigl(~\bar{a}-\beta \cdot  \overline{|r|}~\bigr)^2} = min
~~~~~~~=>~~~~\beta = 6-8 \ln 2 \approx 0.455.
\end{equation}
However, $\sigma_v/\bar{v}$ is not the only criterion, and due to
the low sensitivity of the relative volatility to change in $\beta$, 
we use in this paper the value $\beta=1/2$ and notation $v=a-|r|/2$. 
In what follows we denote $v_\beta=a-0.455 \cdot|r|$.

We note that there is another simple measure of volatility,
comparable in its effectiveness to (\ref{appendix_v_beta}), namely:
\begin{equation}\label{appendix_v_F}
  v_F = \frac{a}{1+r^2/a^2}.
\end{equation}
Although the probability of zero value $a$ for finite duration of
a lag $T$ is vanishingly small, it is still necessary to
define the corresponding value $v_F=0$ for $a=0$. 
Actually, the relations (\ref{appendix_v_beta})
and (\ref{appendix_v_F}) are not analytic functions on $a$ and $r$,
and thus are not governed by the lemma from Appendix B
of \cite{Garman:1980}.

$\bullet$ In addition to the width of distribution, sometimes 
absent or weak dependence on the drift $\mu$ are used as a
criterion. Note that for daily, or shorter, lags $\mu \ll \sigma$;
therefore, this criterion is not that significant. The above
proposed measure of the modified lag range, as well as the
price range itself, depends on $\mu$. However, this dependence is
significantly weaker for $v$ than for the amplitude $a$. If we use the
presentations (\ref{appendix_aver_h_r_mu}),
(\ref{appendix_aver_l_a_mu}), we can write the following 
expression for $v_\beta$:
\begin{equation}
\overline{v}_\beta = (2-\beta)\cdot \sqrt{\frac{2}{\pi}} + \frac{(2-3\beta)\mu^2}{3\sqrt{2\pi}} - \frac{(2-5\beta)\mu^4}{60\sqrt{2\pi}} +...~~~~~
\end{equation}
It can be seen that the factor beside $\mu^2$ for $\beta=1/2$ is
four times smaller than for $\beta=0$ ($v_P=a$). Consequently, the
dependence on $\mu$ is four times weaker as well. When $\beta=2/3$
(denoted $v_{2/3}$ below) the coefficient at $\mu^2$ becomes equal
to zero, and the dependence on $\mu$ is weakening still, although
it disappears completely only for the measure by Rogers and
Satchell.

$\bullet$ Let us now compare the statistical parameters of different
volatility measures shown in Table~\ref{tbl:6}.
\begin{table}[H]
\centering
\small
\caption{Statistical parameters of different volatility estimators, derived analytically (upright) and numerically ({\it italic})}
\label{tbl:6}
\begin{tabular}{l|rrrrrr|rr}
Measure     & $\bar{v}$    & $\overline{v^2}$& $\sigma_v$   & $as$       & $ex$  &  $p_1$ &$\sigma_v/\bar{v}$  &$\sigma_{v^2}/\overline{v^2}$ \\
\hline
$v_P$       &       1.596  &     2.773 &       0.476 &        0.97 &       1.24 &{\it 70.6}&      0.298&      0.638 \\
$v_{RS}$    &  {\it 0.960} &{\it 0.998} & {\it 0.275} & {\it 0.46} & {\it 0.42} &{\it 69.5}& {\it 0.287}&{\it  0.576 }\\
$v_{2/3}$   &       1.064 &      1.217 &       0.292 &  {\it 0.52} & {\it 0.29} &{\it 68.4}&      0.275& {\it  0.557 }\\
$v_{GK}$    &  {\it 0.968} &{\it 0.998} & {\it 0.245} & {\it 0.60} & {\it 0.39} &{\it 68.6}& {\it 0.253}&{\it  0.519 }\\
$v_F$       &  {\it 1.254} &{\it 1.673} & {\it 0.316} & {\it 0.53} & {\it 0.28} &{\it 68.4}& {\it 0.252}&{\it  0.513 }\\
$v$         &       1.197  &     1.523 &       0.300 &       0.53  &      0.26  &{\it 68.2}&      0.251& {\it  0.511 }\\
$v_\beta$   &       1.233  &     1.615 &       0.308 &  {\it 0.55} & {\it 0.29} &{\it 68.3}&      0.250& {\it  0.510 }\\
\end{tabular}
\end{table}
We use italic font to mark the values obtained by Monte Carlo simulation 
for 3.5 million lags, each being a random walk of 1 million ticks. In this
case, for means and volatility an error of order of $\pm 0.002$ is
possible in the last significant digit. The other values
(in upright font) are derived through analytical calculations.

\begin{wrapfigure}{r}{6cm}
\includegraphics{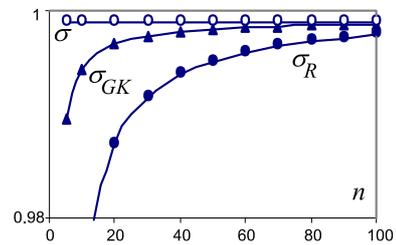}
\caption{Mean values of volatility for different sample sizes $n$ for
several estimators.}\label{fig:33}
\end{wrapfigure}
$\bullet$ For non-stationary data it is often
necessary to conduct the averaging over a relatively small number of
observations $n$. In this case, a bias becomes apparent in quadratic
measures for volatility $\sigma$.
Even if one calculates the classical squared volatility $\sigma^2_R$
by means of the unbiased formula (\ref{appendix_sigma_R_P}), the value
$\sigma_R$ will be biased; indeed, when averaging over large
numbers of samples of size $n$, we have
$<\sigma^2_R>=\sigma^2$, but $<\sqrt{\sigma^2_R}> \neq\sigma$. If
we are  interested in the value of volatility itself rather than its square,
it is better to use linear rather than quadratic measures for
non-stationary data.

To illustrate the effect of drift we provide charts of mean values
of volatility (Fig.~\ref{fig:33}), obtained by averaging a large number
of samples of $n$ values each, for standard definition of
$\sigma_R$ and $\sigma_{RG}=\sqrt{v^2_{RG}}$ measure
(\ref{appendix_sigma_RG}) compared to a linear measure of
$\sigma=(a-|r|/2)\sqrt{2\pi}/3$.

Thus, the measure $v=a-|r|/2$ has a relatively narrow distribution
and consecutively results in smaller error in volatility measurement. 
Simplicity is its obvious advantage, as compared with the measures $v_{RS}$
and $v_{GK}$. In addition, it is unbiased in case of small
sample size, which is significant in examining the effects of
non-stationarity.


\newpage

\bibliographystyle{unsrt}
\bibliography{volatility}

\begin{thebibliography}{10}

\bibitem{Schwert:1989}
G~William Schwert.
\newblock Why does stock market volatility change over time?
\newblock {\em Journal of Finance}, 44(5):1115--53, December 1989.

\bibitem{Shephard:2005}
Neil Shephard, editor.
\newblock {\em Stochastic volatility: selected readings}.
\newblock Oxford Univ. Press, 2005.

\bibitem{Poon:2003}
Ser-Huang Poon and Clive W.~J. Granger.
\newblock Forecasting volatility in financial markets: A review.
\newblock {\em Journal of Economic Literature}, 41(2):478--539, June 2003.

\bibitem{Hsieh:1991}
D.A. Hsieh.
\newblock {Chaos and nonlinear dynamics: application to financial markets}.
\newblock {\em Journal of Finance}, 46(5):1839--1877, 1991.

\bibitem{Watanabe:2008}
M.~Watanabe.
\newblock {Price Volatility and Investor Behavior in an Overlapping Generations
  Model with Information Asymmetry}.
\newblock {\em The Journal of Finance}, 63(1):229--272, 2008.

\bibitem{Battalio:2006}
R.~Battalio and P.~Schultz.
\newblock {Options and the Bubble}.
\newblock {\em Journal of Finance}, 61(5):2071, 2006.

\bibitem{McAleer:2008}
Michael McAleer and Marcelo Medeiros.
\newblock Realized volatility: A review.
\newblock {\em Econometric Reviews}, 27(1-3):10--45, 2008.

\bibitem{Shephard:2008}
Neil Shephard and Torben Andersen.
\newblock Stochastic volatility: Origins and overview.
\newblock Economics Papers 2008-W04, Economics Group, Nuffield College,
  University of Oxford, May 2008.

\bibitem{Broto:2004}
C.~Broto and E.~Ruiz.
\newblock {Estimation methods for stochastic volatility methods: a survey}.
\newblock {\em Journal of Economic Surveys}, 18(5):613--649, 2004.

\bibitem{Andersen:1998}
Torben~G Andersen and Tim Bollerslev.
\newblock Answering the skeptics: Yes, standard volatility models do provide
  accurate forecasts.
\newblock {\em International Economic Review}, 39(4):885--905, November 1998.

\bibitem{Forecasting:2007}
J.Knight and S.~Satchell, editors.
\newblock {\em Forecasting Volatility in the Financial Markets}.
\newblock Elsevier, third edition, 2007.

\bibitem{Engle:2001}
Robert~F Engle and Andrew~J Patton.
\newblock What good is a volatility model.
\newblock {\em Quantitative Finance}, 1(2):237--245, 2001.

\bibitem{Engle:1982}
Robert~F Engle.
\newblock Autoregressive conditional heteroscedasticity with estimates of the
  variance of united kingdom inflation.
\newblock {\em Econometrica}, 50(4):987--1007, July 1982.

\bibitem{Engle:2002}
Robert Engle.
\newblock New frontiers for arch models.
\newblock {\em Journal of Applied Econometrics}, 17(5):425--446, 2002.

\bibitem{Alizadeh:2002}
Sassan Alizadeh, Michael~W. Brandt, and Francis~X. Diebold.
\newblock Range-based estimation of stochastic volatility models.
\newblock {\em Journal of Finance}, 57(3):1047--1091, 06 2002.

\bibitem{Eraker:2004}
Bjørn Eraker.
\newblock Do stock prices and volatility jump? reconciling evidence from spot
  and option prices.
\newblock {\em Journal of Finance}, 59(3):1367--1404, 06 2004.

\bibitem{Canina:1993}
Linda Canina and Stephen Figlewski.
\newblock The informational content of implied volatility.
\newblock {\em Review of Financial Studies}, 6(3):659--81, 1993.

\bibitem{Mikosch:2004}
Thomas Mikosch and C\u{a}t\u{a}lin St\u{a}ric\u{a}.
\newblock Nonstationarities in financial time series, the long-range
  dependence, and the igarch effects.
\newblock {\em The Review of Economics and Statistics}, 86(1):378--390, 01
  2004.

\bibitem{Granger:1999}
Clive W.~J. Granger and Timo Terasvirta.
\newblock A simple nonlinear time series model with misleading linear
  properties.
\newblock {\em Economics Letters}, 62(2):161--165, February 1999.

\bibitem{Diebold:2001}
F.X. Diebold and A.~Inoue.
\newblock {Long memory and regime switching}.
\newblock {\em Journal of Econometrics}, 105(1):131--159, 2001.

\bibitem{Mandelbrot:1963}
Benoit Mandelbrot.
\newblock The variation of certain speculative prices.
\newblock {\em Journal of Business}, 36:394, 1963.

\bibitem{Jondeau:2007}
Eric Jondeau, Ser-Huang Poon, and Michael Rockinger.
\newblock {\em Financial Modelling under non-Gaussian Distributions}.
\newblock Springer, 2007.

\bibitem{Fama:1965}
Eugene~F. Fama.
\newblock The behavior of stock-market prices.
\newblock {\em The Journal of Business}, 38(1):34--105, 1965.

\bibitem{Parkinson:1980}
Michael Parkinson.
\newblock The extreme value method for estimating the variance of the rate of
  return.
\newblock {\em Journal of Business}, 53(1):61--65, January 1980.

\bibitem{Garman:1980}
Mark~B Garman and Michael~J Klass.
\newblock On the estimation of security price volatilities from historical
  data.
\newblock {\em Journal of Business}, 53(1):67--78, January 1980.

\bibitem{Rogers:1991}
L.C.G. Rogers and S.E. Satchell.
\newblock {Estimating variance from high, low and closing prices.}
\newblock {\em Annals of Applied Probability}, 1(4):504--512, 1991.

\bibitem{Rogers:2008}
L.~C.~G. Rogers and Fanyin Zhou.
\newblock Estimating correlation from high, low, opening and closing prices.
\newblock {\em ANNALS OF APPLIED PROBABILITY}, 18:813, 2008.

\bibitem{Yang:2000}
Dennis Yang and Qiang Zhang.
\newblock Drift-independent volatility estimation based on high, low, open, and
  close prices.
\newblock {\em Journal of Business}, 73(3):477--91, July 2000.

\bibitem{Barndorff:2002}
Ole~E. Barndorff-Nielsen and Shephard.
\newblock Econometric analysis of realized volatility and its use in estimating
  stochastic volatility models.
\newblock {\em Journal Of The Royal Statistical Society Series B},
  64(2):253--280, 2002.

\bibitem{Biais:2005}
Bruno Biais, Larry Glosten, and Chester Spatt.
\newblock Market microstructure: A survey of microfoundations, empirical
  results, and policy implications.
\newblock {\em Journal of Financial Markets}, 8(2):217--264, May 2005.

\bibitem{Andersen:2003}
Torben~G. Andersen, Tim Bollerslev, Francis~X. Diebold, and Paul Labys.
\newblock Modeling and forecasting realized volatility.
\newblock {\em Econometrica}, 71(2):579--625, March 2003.

\bibitem{Madhavan:2000}
Ananth Madhavan.
\newblock Market microstructure: A survey.
\newblock {\em Journal of Financial Markets}, 3(3):205--258, August 2000.

\bibitem{Bandi:2006}
Federico~M. Bandi and Jeffrey~R. Russell.
\newblock Separating microstructure noise from volatility.
\newblock {\em Journal of Financial Economics}, 79(3):655--692, March 2006.

\bibitem{Bollerslev:1993}
Tim Bollerslev and Ian Domowitz.
\newblock Trading patterns and prices in the interbank foreign exchange market.
\newblock {\em Journal of Finance}, 48(4):1421--43, September 1993.

\bibitem{Cont:2001}
R.~Cont.
\newblock Empirical properties of asset returns: stylized facts and statistical
  issues.
\newblock {\em Quantitative Finance}, 1(2):223--236, 2001.

\bibitem{Ding:1993}
Zhuanxin Ding, Clive W.~J. Granger, and Robert~F. Engle.
\newblock A long memory property of stock market returns and a new model.
\newblock {\em Journal of Empirical Finance}, 1(1):83--106, June 1993.

\bibitem{Breidt:1998}
F.~Jay Breidt, Nuno Crato, and Pedro de~Lima.
\newblock The detection and estimation of long memory in stochastic volatility.
\newblock {\em Journal of Econometrics}, 83(1-2):325--348, 1998.

\bibitem{Hodrick:1997}
Robert~J Hodrick and Edward~C Prescott.
\newblock Postwar u.s. business cycles: An empirical investigation.
\newblock {\em Journal of Money, Credit and Banking}, 29(1):1--16, February
  1997.

\bibitem{Barndorff:2001}
Ole~E. Barndorff-Nielsen and Neil Shephard.
\newblock Non-gaussian ornstein-uhlenbeck-based models and some of their uses
  in financial economics.
\newblock {\em Journal Of The Royal Statistical Society Series B},
  63(2):167--241, 2001.

\bibitem{Feller:1951}
William Feller.
\newblock The asymptotic distribution of the range of sums of independent
  random variables.
\newblock {\em The Annals of Mathematical Statistics}, 22:427--432, September
  1951.

\bibitem{Borodin:2000}
A.N. Borodin and P.~Salminen.
\newblock {\em Handbook of Brownian Motion - Facts and Formulae}.
\newblock Basel: Birkhauser, 2000.

\end{thebibliography}

\end{document}